# Design and Planning of Flexible Mobile Micro-Grids Using Deep Reinforcement Learning

Cesare Caputo, Michel-Alexandre Cardin,

Pudong Ge, Fei Teng, Anna Korre, Ehecatl Antonio del Rio Chanona

*Imperial College London*

## Abstract

Ongoing risks from climate change have significantly impacted the livelihood of global nomadic communities, and are likely to lead to increased migratory movements in coming years. As a result, mobility considerations are becoming increasingly important in energy systems planning, particularly to achieve energy access in developing countries. Advanced "Plug and Play" control strategies have been recently developed with such a decentralized framework in mind, more easily allowing for the interconnection of nomadic communities, both to each other and to the main grid. In light of the above, the design and planning strategy of a mobile multi-energy supply system for a nomadic community is investigated in this work. Motivated by the scale and dimensionality of the associated uncertainties, impacting all major design and decision variables over the 30-year planning horizon, Deep Reinforcement Learning is implemented for the design and planning problem tackled. DRL based solutions are benchmarked against several rigid baseline design options to compare expected performance under uncertainty. The results on a case study for ger communities in Mongolia suggest that mobile nomadic energy systems can be both technically and economically feasible, particularly when considering flexibility, although the degree of spatial dispersion among households is an important limiting factor. Key economic, sustainability and resilience indicators such as Cost, Equivalent Emissions and Total Unmet Load are measured, suggesting potential improvements compared to available baselines of up to 25%, 67% and 76%, respectively. Finally, the decomposition of values of flexibility and plug and play operation is presented using a variation of real options theory, with important implications for both nomadic communities and policymakers.

## **Nomenclature Table**

| **Acronyms** | | **Indices and Sets** | |
|---|---|---|---|
| ACER | Actor Critic with Prioritized Experience Replay | $t \in T$ | Index and set of *general* time steps |
| BD1/2/3/4 | Baseline Rigid Design 1/2/3/4 | $m \in M$ | Index and set of *monthly* decision time steps |
| BESS | Battery Energy Storage System | $h \in H$ | Index and set of operational simulation time steps |
| DGs | Distributed Generators | $s \in S$ | Index and set of scenarios |
| DNNs | Deep Neural Networks | $i \in I$ | Index and set of technological components |
| DRL | Deep Reinforcement Learning | $x \in X$ | Index and set of GBM modeled variables |
| EH | Electric Heating/Heater | $n \in N$ | Index and set of nanogrids/nodes in PP microgrid |
| ENPC | Expected Net Present Cost | **Model Variables and Parameters** | |
| EoS | Economies of Scale | $T$ | Evaluation horizon (months) |
| FD | Flexible Design | $\xi$ | Uncertainty vector |
| FND | Flexible Nanogrid Design | $\theta$ | Installed nominal capacity (kW or kWh) |
| GA | Genetic Algorithm | $\lambda$ | Financial discount rate |
| GBM | Geometric Brownian Motion | $\mathcal{R}$ | Revenue Function |
| LFS | Load Following Strategy | $\mathcal{C}$ | Cost Function |
| MDP | Markov Decision Process | $\mathcal{H}$ | Expansion Cost Function |
| ML | Machine Learning | $CAPEX$ | Initial Capital Investment |

| | | | | |
|---|---|---|---|---|
| MS | Mobility Scenario | $NCF$ | Net cash flow (USD) |
| MSSP | Multi-Stage Stochastic Programming | $NPC$ | Net present cost (USD) |
| NPC | Net Present Cost | $ENPC_{es}$ | Energy system ENPC (USD) |
| PP | Plug and Play | $C$ | Energy system cost (USD) |
| REAP | Rural Electricity Access Project | $s$ | State vector |
| RES | Renewable Energy System | $a$ | Action vector |
| SDG | Sustainable Development Goal | $A$ | Action space |
| SHS | Small Home Solar | $R$ | Reward |
| SWS | Small Wind Systems | $\gamma$ | DRL discount rate |
| TUL | Total Unmet Load | $\pi$ | DRL agent policy |
| UB | Ulaanbaatar | $VoF$ | Value of flexibility |

**Model Variables and Parameters (continued)**

| | | | |
|---|---|---|---|
| $P_{ul,es}$ | Unmet total load for *PP* microgrid (kWh) | $EV$ | Expected Value |
| $P_{dist,es}$ | Total distributed electricity in *PP* microgrid (kWh) | $VoPP$ | Value of Plug and Play Operation |
| $\theta_{PP,es}$ | Total *PP microgrid* distribution capacity limits (kW) | $\mu$ | Drift |
| $P_{dist,es,max}$ | Constraint on maximum *PP microgrid* distribution (% of ED) | $\sigma$ | Volatility |
| $P_{dist}$ | Active power distribution requirements per *nanogrid* (kW) | $W$ | Weiner process stochastic deviation |
| $L_{dist}$ | Distance distribution requirements per *nanogrid* (kW) | $N_n$ | Number of community dwellings |
| $R_{dist}$ | Distribution resistance requirements per *nanogrid* ($\Omega$) | $r_{cluster}$ | Nomadic community cluster radius (km) |
| $V_{dist}$ | Voltage distribution requirements per *nanogrid* (V) | $L_{cb}$ | Cabling length (km) |
| $I_{dist}$ | Current distribution requirements per *nanogrid* (amp) | $P_{wind}$ | Active Wind power (kW) |
| $\alpha_{cb}$ | Material temperature coefficient | $P_{pv}$ | Active PV power (kW) |
| $P_{dl,es}$ | Active power distribution losses for *PP microgrid* (kW) | $P_{NEL}$ | Net electricity load (kW) |
| $DL$ | Total distribution loss % | $P_{NHL}$ | Net heating load (kW) |
| $P_{grid}$ | Active exchanged with main grid for *PP microgrid* (kW) | $ED$ | Electricity demand (kW) |
| $\theta_{grid}$ | Total *main grid* distribution capacity limits (kW) | $HD$ | Heating demand (kW) |
| $\rho_{grid}$ | Binary variable on grid availability | $P_{res}$ | RES active power (kW) |
| $P_{uhl,es,ppp}$ | Unmet heating load after *PP microgrid* distribution (kW) | $P_{eh,max}$ | Maximum EH capacity (kW) |
| $P_{res,exc,pp}$ | Excess electricity after *PP microgrid* distribution (kW) | $\eta_{inv}$ | Inverter roundtrip efficiency (%) |
| $P_{coal,es}$ | Power coal requirements for *PP microgrid* (kW) | $P_{ba,in,max}$ | Maximum battery charging capacity (kW) |
| $M_{coal}$ | Coal mass requirements for *PP microgrid* (kg) | $P_{ba,out,max}$ | Maximum battery discharging capacity (kW) |
| $\mu_{local\ coal}$ | Local lignite coal energy density | $E_{ba}$ | Battery storage level (kWh) |
| $\eta_{coal\ stove}$ | Coal stove net efficiency (%) | $E_{ba,min}$ | Minimum battery storage level (kWh) |
| $CO_{2\ eq.emissions}$ | Equivalent $CO_2$ emissions generated (tonnes) | $E_{ba,max}$ | Maximum battery storage level (kWh) |
| $EF$ | Weighted average emission factor | $\eta_{batt}$ | Battery roundtrip efficiency (%) |
| $C_{inv}$ | Total investment costs (USD) | $P_{res,PL}$ | Available RES capacity post meeting nanogrid load (kW) |
| $C_{opex}$ | Operational costs (USD) | $N_{deficit}$ | Nanogrid units with electricity deficit |
| $C_{grid}$ | Grid active power exchange costs or revenue (USD) | $N_{surplus}$ | Nanogrid units with electricity surplus |
| $C_{CO2}$ | Carbon emissions equivalent costs or revenue (USD) | $P_{eg,es}$ | Excess *PP microgrid* electricity generation (kWh) |
| $C_{es,max}$ | Budget constraint (USD) | $L$ | Functional operational life (years) |
| $CC$ | Communication costs (USD) | $SV$ | Salvage value (USD) |
| $EC$ | Expansion costs (USD) | $\theta_{sv}$ | Abandoned capacity (kW) |
| $RC$ | Replacement costs (USD) | $k$ | Depreciation factor |
| $IC$ | Infrastructure costs (USD) | $C_{coal}$ | Total coal costs (USD) |
| $PEC$ | Power electronics costs (USD) | $c_{coal}$ | Coal cost factor (USD/kg) |
| $c_{cc}$ | Communication costs factor (USD/kWh) | $C_{om}$ | Total O&M costs (USD) |
| $c_{ic}$ | Cabling costs factor (USD/ km) | $c_{om}$ | O&M cost factor (USD/kW) |
| $c_{pe}$ | Power electronics cost factor (USD/kW) | $C_{ul}$ | Total unmet load costs (USD) |
| $K$ | Investment constant in EoS model | $c_{ul}$ | Unmet load cost factor (USD/kWh) |
| $\alpha$ | EoS factor in EoS model | | |
| $\varepsilon_{pe}$ | Power electronics requirement ratio | $C_{grid}$ | Total grid interaction costs (USD) |
| $sf_{pe}$ | Safety factor | $c_{grid}$ | Grid tariff (USD/kWh) |
| $M_{batt}$ | Battery total weight (kg) | $p_{grid}$ | Grid RES subsidy (USD/kWh) |
| $M_{pe}$ | Power electronics total weight (kg) | $M_{RES}$ | RES total weight (kg) |
| $M_{cb}$ | Distribution cabling total weight (kg) | | |

# 1 Introduction
## 1.1 Climate Change and Migration

The climate crisis experienced in recent years is inducing significant changes in temperature, precipitation, and extreme weather event patterns [1]. This is impacting the livelihood of nomadic communities, as it is highly dependent on local environmental conditions in areas unsuitable for sedentary agriculture [2]. As such, most of them are dependent on following an approximately fixed seasonal pattern of movements to sustain their agricultural activities, transporting some form of mobile dwelling. It is estimated that there are currently a least 30-40 million rural nomads, primarily found across Sub-Saharan Africa, South-Asia, and Central America [3]. On the other hand, areas where sedentary agriculture can be sustainably practiced are rapidly decreasing, leading to climate-change induced mobility worldwide [4]. Recent models suggest that around 143 million more people will be displaced by 2050, likely impacting vulnerable communities the most [5, 6].

Climate driven migration is also recognized as one of the major causes of urbanisation worldwide [7]. Urban migrants are normally composed primarily of unskilled workers associated with the informal sector, where the lack of proper steady housing has limited their integration into the formal sector. They typically create informal settlements on the outskirts of major cities, housed in small tent-like structures. As such, they are normally unable to access services such as electricity or district heating. The resulting energy poverty impacts the livelihood of urban migrants significantly, having to spend long periods of time and a high proportion of their disposable income in the search for more rudimentary power sources, such as dung or biomass [8]. Furthermore, it was estimated that globally there are at least 208 million urban migrants completely lacking access to energy, primarily due to infrastructure limitations or lack of access to the formal government sector [7, 9].

While the UN Sustainable Development Goals (SDGs) do not explicitly link climate change and migration, it is becoming increasingly important to develop policies and technologies which help mitigate the impacts of climate induced mobility [10]. Energy access is very complex in this context, as the current infrastructure is not compatible with such mass migratory movements, particularly in lower income countries [11]. In order to address SDG 7 (*Affordable and Clean Energy for All*) it is, therefore, imperative to develop energy system solutions which are compatible with a migratory lifestyle, as opposed to the "static" approach used in energy systems planning thus far [12].

## 1.2 Need for Plug and Play Based Flexible and Sustainable Energy Systems

### 1.2.1 Overview

Given the current high costs for national grid extensions, mini-grid and micro-grid solutions are expected to be the main driver of electrification for almost half of all households currently lacking energy access [13]. The rapid cost decline for renewable energy systems (RES) over the past few years has already increased their rate of deployment, providing electricity to 140 million people as of 2020 [9]. Similarly, there are ongoing trends towards more distributed generators (DG) as RES are increasingly installed closer to their respective loads, helping reduce transmission power losses. The uncertainties and intermittence of RES generation, however, lead to several technical challenges for their integration into the main grid. At higher levels of RES penetration in particular, maintaining operational parameters such as voltage and frequency within their allowable limits can become increasingly problematic [14].

Nomadism represents a more extreme example of the above issues, as both loads, and generation are not only distributed but also changing over different temporal and spatial scales. Simultaneously, remote working policies implemented as a result of the COVID-19 pandemic have led to a global increase in "digital nomads" [15]. As a lot of these individuals move around in campers or recreational vehicles, they contribute to increasingly decentralized and spatially unpredictable electricity generation and demand. Based on the above and the trends mentioned in 1.1, there is a growing need to develop flexible and mobile energy systems - which are able to switch operation between a centralized and decentralized manner to accommodate both rural and urban migration.

Recent advancements in distributed control algorithms have enabled the consideration of mobile and decentralized energy system solutions in the context with a migratory lifestyle [16, 17]. In most regions currently practicing nomadism, in fact, family clusters tend to be relatively close to each other, following similar seasonal migration patterns [3]. This creates the possibility for temporary interconnection among them to form a mobile herder household energy supply system (i.e. *micro-grid*), or allowing flexible plug-in operations with the main grid. This sort of plug and play (PP) operation can be highly advantageous as it allows the smoothing out of load/generation imbalances among various households, with potentially different electricity generation and storage capacities. Plug-in operations with the main grid, on the other hand, could be beneficial by providing an additional energy supply to nomadic communities when available, while helping improve main grid stability through a more distributed storage capacity [18]. Additionally, market engagement opportunities may be generated, potentially helping to reduce the total electrification costs for nomadic communities.

Moreover, PP holds potential to accelerate the decarbonization of heating supply systems for nomadic communities living in colder regions. Their traditional heating systems rely on inefficient combustion of fossil fuels and could be replaced by electric-heating (EH) technologies if a reliable and sufficient level of sustainable electrification can be achieved. This may be particularly attractive if implemented at a community rather than single household level, as a higher degree of modularity can be integrated in system capacity evolution strategies [7]. Currently, however, all PP studies have focused on fine timestep resolution control operations to guarantee network reliability. Furthermore, none have analysed it through a multi-energy vector perspective to accelerate the transition to sustainable heating. Given the potential long-term implications of PP as climate induced migration increases, a novel modelling and planning framework which integrates mobility considerations is needed to quantify its long-term impact, and develop policy recommendations for both rural and urban migrants.

### 1.2.2 Plug and Play (PP) Operation

PP operation can be allowed without control law changes in the power-electronic-level devices, guaranteeing system stability and synchronisation (in terms of frequency and voltage), as detailed in 2.1.3. As such, it is compatible with operational control strategies including predictive Machine Learning (ML) methods, advanced control algorithms, or heuristic-based approaches [14, 19-23].

Nonetheless, recent studies suggest that for smaller scale systems with no dispatchable generator, a simple Load-Following Strategy (LFS) is likely to result in similar conclusions on initial design capacities and total costs as more complex methods [23-25]. Its logic prioritises renewable generation, followed by battery energy storage and grid interactions, if available, to meet any remaining load/demand – adopting simple heuristics to operate the system. Accordingly, it is well suited to the technological and communication capacity limitations associated with a nomadic mobile energy supply system. Based on the more limited control opportunities available, a flexible capacity planning strategy under uncertainty is, therefore, more likely to impact the distribution of total system costs (as well as other performance metrics) obtained through long-term PP operation.

### 1.2.3 Energy System Planning and Role of Flexibility

General micro-grid planning over a long horizon is a complex problem that has been extensively investigated in the literature, and is closely related to rural electrification objectives [26-29]. The aim is typically minimisation of Net Present Cost (NPC) although there are a few works which have more closely looked into sustainability [30, 31] and resilience [17, 32, 33] considerations too. Traditionally, to help reduce the computational expense of the optimisation, planning problems can be simplified to single-year approaches, where the performance of the system over the entire planning horizon is determined based on the simulation of only one "representative" year [34]. While the computational simplifications are substantial, this kind of approach does not account for load growth or corresponding energy system evolution. These are key elements when considering rural electrification objectives, where load growth can be particularly uncertain, thus multi-year approaches are more applicable in the targeted context [35, 36].

The majority of these approaches are based on either Mixed-Integer-Linear Programming (MILP), Linear Programming (LP), Multi-Stage Stochastic Programming (MSSP), or evolutionary methods such as the genetic algorithm (GA) [28]. In most works, however, to address the non-linearities of certain model components and problem dimensionality, significant simplifications tend to be made – particularly in terms of uncertainty recognition and quantification. These may also include considering only a specific component of the system (i.e., battery in [37] for example), static energy system capacities through project duration or only using a few representative days for each year [28, 29]. The manuscript in [38], for example, implemented a multi-year approach for the planning of a microgrid with no dispatchable generator, but did not account for any stochastic elements nor a complete planning horizon. Similarly, other works focused on rural electrification, while modelling load growth stochastically, do not account for uncertainty on energy generation over the project duration [39]. Typically, only one or two sources of uncertainty are considered, and their realizations limited to a predefined set of scenarios. Furthermore, the majority of these approaches limit decision-making flexibility options through the project lifetime, primarily to maintain computational tractability. Generally, in fact, only one level of expansion is considered for each technology, and expansion decisions are only made based on scenario realizations, with limited generalizability [24, 25, 40].

The planning and design of a nomadic energy supply system, however, must balance delicate trade-offs between mobility, security, efficiency, sustainability, and economic feasibility in light of the extensive associated operational uncertainties. Flexibility is a potential value enhancing paradigm to help address some of the above-mentioned limitations of common energy system planning approaches [41, 42]. *Flexibility in Design* explicitly considers decision-making possibilities over a systems lifetime, allowing for more cost-effective adaptability, sustainability, and resilience under uncertainty. This is highly applicable for evolvability in complex energy systems projects. This field aims to produce computational tools to support early conceptual design phases as part of a holistic framework, with primary applications across uncertainty modelling, concept generation, and design space exploration [41]. These methods help produce modularised designs with decreased exposure to downside risks while capitalizing on upside opportunities, shifting the distribution of performance as compared to standard approaches. Improvements typically range from 10-30%, although several energy system application studies suggest even greater expected performance enhancements may be achievable, depending on the problem at hand [43-46]. Planning a PP based mobile energy system through a *Flexibility in Design* approach can, therefore, likely help mitigate investment risk and fiscal burdens incurred by the targeted low-income nomadic communities under uncertainty.

Given the degree of uncertainty these rural isolated energy system planning problems normally face, it is essential to consider a greater level of decision-making flexibility to enable adaptability to different encountered conditions. Deep Reinforcement Learning (DRL), while widely investigated and proven for applications in energy systems management and control [47], has yet to be directly applied to multi-year microgrid planning and design under several sources of uncertainty. The closest work would be that by [48], however DRL was only used by the authors for the dispatch problem, and components were sized through evolutionary methods. Additionally, energy system capacity planning or decision-making flexibility under uncertainty were not considered. An important motivation for this work is, therefore, that a data-driven based approach may help in the analysis of potential value and costs of system flexibility in different scenarios, by allowing an expanded design, strategy, and uncertainty space as compared to alternative state of the art methods such as MILP or MSSP [49].

### 1.3 Data-Driven Energy Systems Design for Flexibility

Standard methods for *Flexibility in Design*, however, present many of the same computational limitations as the planning approaches mentioned in the former section. Typically, they are also based on an MSSP or decision tree formulation over a long time horizon [50]. At each stage, the possible actions represent system adaptability options, and the optimal solution quickly becomes intractable if several uncertainty sources or flexibility strategies are considered [51]. Path independence must be assumed, which is unrealistic in the context of a nomadic energy system. The solution procedure has also been criticized, as one that is difficult to understand and implement in practice for decision makers [52]. Decision rules help address some of these issues, by mapping system evolution strategies to physical design variables in a way which is intuitive for decision makers [41]. Although they have been shown to estimate similar values of flexibility as more complex alternatives, strategies are still limited to generic real options or domain specific ones.

A data-driven approach to *Flexibility in Design* may help address some of these limitations, by more systematically evaluating the potential decision rule space, as shown recently by Caputo and Cardin [49]. Their results on an example energy systems design and planning case study suggest that Deep Reinforcement Learning (DRL) can enhance solution expected value under uncertainty, leading to several potentially interesting applications. DRL is a subset of ML defined by a sequential decision making process, and is best known for its state of the art results on game-like environments [53, 54]. In recent years, however, it has been successfully implemented in several engineering domains with significant variations in action and state spaces, as well as objectives [53, 55-59]. In the energy systems field, it has been extensively investigated for management and control [47], but it has yet to be directly implemented on micro-grid design and capacity planning for rural electrification.

Generally, the aim would be to iteratively find the best adaptability strategies under a wide range of scenarios by balancing exploration and exploitation in the learning process. Deep Neural Networks (DNNs) can be used to accurately approximate a complex decision-making policy and corresponding value of each flexibility strategy over time. This allows for the tractable integration of several sources of uncertainties and mobility considerations associated with a mobile energy supply system. The design and uncertainty space relevant to an energy system planning problem, it follows, can be significantly expanded as opposed to analytical methods such as MSSP [49]. Consequently, a data-driven approach may also help identify novel, unintuitive and more dynamic adaptation strategies outside of the standard generic real options framework. This is particularly valuable in the case of a unique, complex, and highly uncertain context such as the one tackled in this work, where there is limited domain expertise. As a result, energy system designs and flexibility strategies developed through a data-driven approach are likely more compatible with (and representative of) the range of real-world conditions potentially faced by mobile energy systems in the future.

## 1.4 Contributions

In light of the pressing issues discussed above, this paper presents the design and planning of a mobile multi-energy supply system which is compatible with a nomadic lifestyle, seeking to integrate both electrification and heating supply decarbonisation objectives. To address deficiencies of common "static" energy planning approaches in this context, a new modelling framework is proposed where the spatial dispersion (and network configuration) among nomadic dwellings varies based on seasonal requirements, as well as main grid availability. In order to tackle the resulting highly complex and dimensional stochastic optimisation problem, a novel DRL based data-driven approach is proposed to avoid the simplifications and compromises associated with typical planning methods mentioned in 1.2.3. The primary contributions are summarized below.

1. Development of a novel modelling framework for the design and planning of mobile multi-energy supply systems, focused on achieving SDG7 for global nomadic communities, but applicable to general trends in increasing decentralization of energy systems. This represents the first study integrating considerations of seasonal centralized to decentralized operation over a long planning horizon, enabled by PP operation. To the authors knowledge, the novel approach proposed is the first allowing for changing network topologies and/or spatial dispersion over time, while capturing the fiscal and mobility constraints faced by a typical targeted nomadic community. The framework is developed to be generic and easily scalable to the design and capacity planning optimisation of different communities wishing to operate a mobile energy supply system.
2. Implementation of a new DRL based approach optimising long-term flexibility strategies for nomadic communities, marking the first time DRL is used in an energy access planning context. Several sources of uncertainty are, as a result, able to be tractably integrated, including on energy generation and demand, spatial dispersion, and main grid availability. This approach enables for the first time the consideration of monthly adaptability strategies, capturing the interactions between mobility and seasonality on optimal system design. Resulting policies can be highly dynamic, and are more generalizable than those obtained using only a few select scenarios. Results are benchmarked against baseline rigid designs with better economies of scale to estimate the values of decision-making flexibility in the context of a migratory lifestyle across economic, sustainability and resilience metrics.
3. Multi-objective assessment of value of PP operation and "breakeven" spatial dispersion point to determine investment guidelines and policy insights to accelerate nomadic energy access. This is performed using a variation of real-options theory and represents the first study focusing on long term planning of mobile PP enabled energy systems enabled. As such, it yields important quantitative results on its practical feasibility and global impact potential.

The novel modelling framework and methodology is exemplified through a case study based on Mongolian herder communities, developed in collaboration with local energy sector stakeholders. The general approach, however, may be easily extended to other nomadic communities worldwide, or to the overall needs to better understand the feasibility of mobile energy systems which switch periodically between centralised and decentralised operation to increase network reliability.

The rest of the paper is organized as follows. Section 2 details the novel modelling framework developed to account for mobility. Section 2.8 introduces the DRL based design and planning optimisation methodology, as well as the benchmark rigid stochastic capacity planning model. Section 4 gives a more in-depth overview of the application case study, while Section 1 summarises the different optimisation parameters implemented. Finally, Section 5 presents the results of the study, accompanied by discussions and conclusions for future work in Sections 6 and 7, respectively.

# 2 Mobile Micro-Grid Model Formulation

## 2.1 Problem Statement

### 2.1.1 Introduction

This section describes the formulation of the model for the planning of mobile-multi energy networks for nomadic communities over a long-term horizon. The integrated model accounts for changing and uncertain movement patterns over time, micro-grid distribution losses, variation in renewable energy output, and different heating generation sources. Furthermore, it considers the possibility of connection/ market engagement with the main distribution grid when available, allowing switching between centralized and decentralized operation based on seasonality and user requirements. Hourly resolution is used for power flow calculations to tractably approximate generation, distribution, and losses at the system level. On the other hand, monthly resolution is implemented for cost calculations within the flexible capacity planning model. This allows for adaptation strategies based on changing climatic conditions and intra-year seasonality, while considering the impact of long-term realisations of uncertainty on mobile micro-grid performance.

Two different mobility scenarios (MS) are thus developed to reflect the diversity in movement pattern. MS1 depicts operation of the micro-grid in the rural or decentralized setting, typically in the summer months for nomadic communities as they migrate to maintain their livestock. MS2, on the other hand, represents operation in a centralized setting, or when connection to the main grid is available, typically in winter months for nomads - when pasture productivity is lowest.

### 2.1.2 Objectives

Two primary alternatives are compared based on the model presented in the following sections for a community operated mobile micro-grid. Firstly, an inflexible baseline is considered where all capacity is installed at project start and no adaptation strategies in response to uncertainty are possible. Secondly, a flexible design exercising modular deployment of different technologies over time based on realised operating conditions is developed. The baseline rigid design presents lower unit costs due to economies of scale and reduced transport charges, while the flexible design allows for the deployment of resources only if and when needed. The objective for both designs is minimization of Expected Net Present Costs ($ENPC_{es}$) for the energy system ($es$) over 30 years ($T$).

Nonetheless, this work is also motivated by the UN SDGs and associated sustainability (i.e., Scope 1 and 2 emissions) and reliability energy access objectives. Emissions are thus found from the direct combustion of coal for heating and environmental footprint of the electricity purchased from the main distribution grid, respectively [60]. Additionally, 30-year horizon total unmet load (TUL) is used as a proxy for estimating system resilience and energy security, considering at a higher level the impact of demand and generation uncertainty. It should be noted that these metrics are not explicitly optimised, but are used for the qualitative evaluation of design alternatives in 5.3.

### 2.1.3 PP Control and Operation

A core enabler of this kind of analysis and system design methodology is the novel potential for PP operation, one of the key functionalities benefitting from the high-penetrated power electronic devices, e.g., inverter. PP operation enables the mobility of energy sources by implementing intelligent algorithms to guarantee operational stability, which is essential for compatibility with a nomadic or migratory lifestyle [17, 61]. This creates an opportunity for economic, sustainability and energy security benefits, even in the rigid baseline design, as further discussed in 5.4. The concept of PP operations only requires real-time power-electronic level control design that has been widely investigated in [62, 63]. The control objective within the micro-grid system is to guarantee that frequency and voltage converge to their reference values (i.e., frequency of 50 Hz and voltage of 220√2 V) to maintain network stability. This is achieved through active power sharing across DGs connected to the microgrid energy supply system. The images on the left of Figure 1 demonstrate the changing microgrid configurations and respective estimated time resolutions for a typical operational scenario encountered. Starting from normal conditions (1), one of the DGs is disconnected in (2) as a family decides to migrate, and it eventually reconnects to the microgrid in (3). The right sub-plots (a)-(d) in Figure 1 demonstrate that PP operation is able to maintain appropriate values for frequency $f(Hz)$, active power ratio $mP$, active power $P(kW)$, and voltage $V_{mag}(V)$, respectively, across scenarios (1)-(3). Over the simulated 5 second horizon, it should be noted, some minor transients are still recorded at a fine resolution due to the changing system loads. Nonetheless, the demonstration results suggest PP operation should be an effective and reliable solution to allow temporary interconnection among nomadic community members.

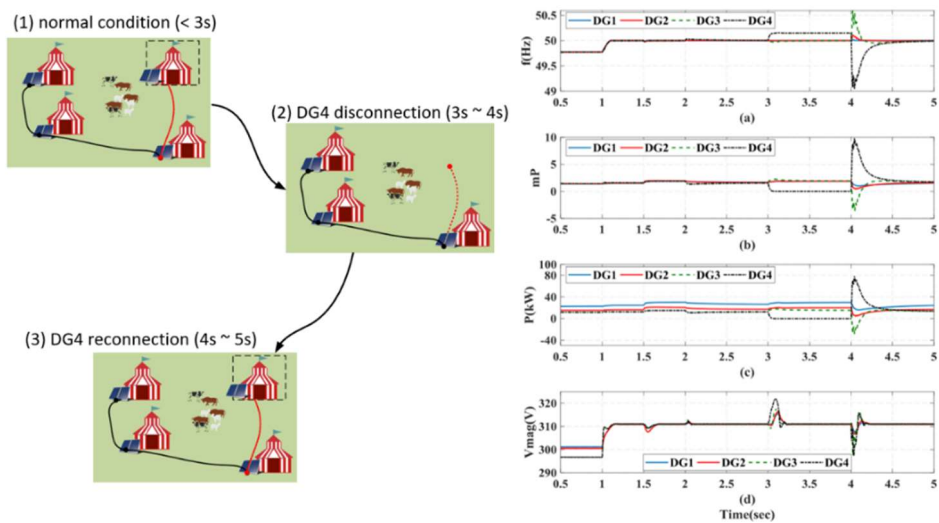

*Figure 1: PP operation demonstration including typical (1) ~ (3) operational scenarios (left) and performance evaluation of frequency, active power ratio, active power, and voltage, in (a)-(d), respectively, over the same time horizon (right).*

## 2.2   Nanogrid and Network Configuration

The architecture of the energy system proposed in this paper is a distributed form of mobile micro-grid. A single herder family *nanogrid (i.e. sub-microgrid) unit* ($N_n$) which moves and settles very close to each other is used as a building block for the simulation, allowing for improved scalability and compatibility with PP operation [64]. The term nanogrid is justified as each herder family unit is normally expected to be able to produce, store and distribute electricity in fully self-sustaining or islanded manner- as sometimes necessary during migration. Through interconnection with other community members, however, the aim is to reduce the total system cost by more efficiently managing supply and demand. Figure 2 graphically captures the network configuration, both at the level of an individual herder family nanogrid unit ($N_n$-left) and the PP mobile micro-grid (right).

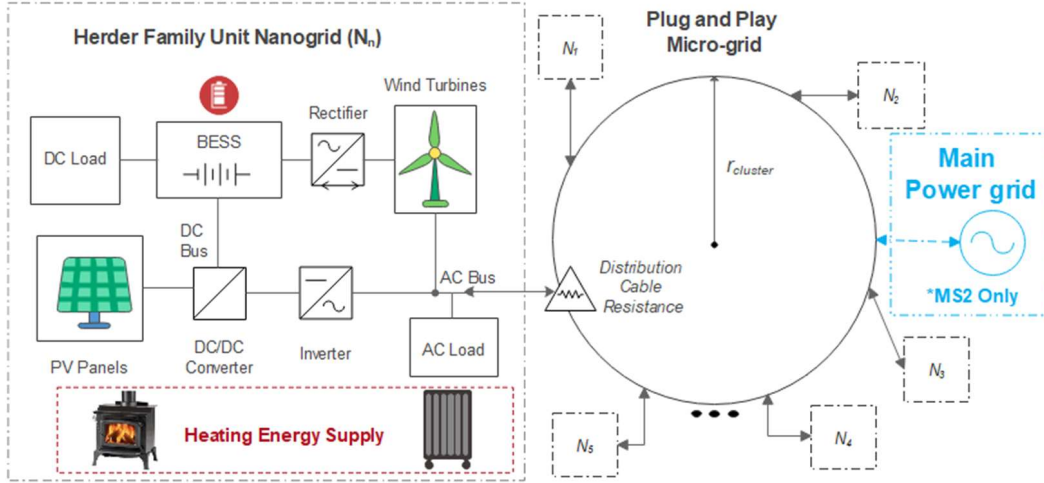

*Figure 2: Multi energy vector block diagram of herder nanogrid (left) and PP microgrid (right)*

The typical nanogrid unit consists of energy generating units – in this case solar PV and wind – as well as a storage system (BESS). While both DC and AC loads may be present in the system (more details in 4.4), interactions among nodes connected through the PP configuration is designed to be 3-phase AC, thus converters and inverters are included where appropriate. Heating demand may be present during the winter months which must be fully met using either a traditional coal stove (left) or electric heater (right). Moreover, it is assumed for it to be possible for a group of nomads, perhaps from the same extended family or community, to operate as a "cluster". The cluster radius variable $r_{cluster}$ is thus defined as the smallest possible circle which could encompass them all – as shown in Figure 2. It is estimated that this is not likely to vary significantly *within* a single season, although there may be important *year-to-year* differences based on the impacts of climate change. Stochastic realizations for a single month $m$ are thus obtained proportionally to average seasonal values at project start, directly impacting the distance ${L_{cb}}^t$ required to connect the mobile energy system, as shown in (1).

$$L_{cb}{}^t = 2\pi r_{cluster}{}^t \tag{1}$$

The different nodes are assumed to be configured in a fixed, ring-like interconnection, evenly spaced across the outer circumference obtained from $r_{cluster}$. This configuration was previously reported to optimise distribution efficiency under a similar decentralised setting as depicted above [14, 19, 21, 65, 66]. The proposed approach can, therefore, account for changing migration distances over time, while also integrating social considerations on spacing among family units. Conditional connection to the main grid during MS2 is approximated as an additional node in the network (blue in Figure 2).

## 2.3 Renewable Energy Generation

The renewables.ninja platform [67] is used for the simulation of wind ($P_{wind}^{n,t}$) and solar PV ($P_{pv}^{n,t}$) intermittent generation for each nanogrid $n$ at an hourly resolution $t$, based on locations representative of each MS. The closest matches in terms of technical performance parameters were used since the small-scale systems considered are not available for direct simulation within the platform. More specifically, Solar PV generation is found through the Global Solar Energy Estimator, with default tilt and azimuth (β = 35°, φ = 180°), an assumed additional system loss fraction of 10%, and no tracking mechanism implemented [68]. Small scale wind generation is computed through the Virtual Wind Farm model and historical wind speeds obtained from NASA's MERRA-2 dataset, implementing bias-corrected reanalysis to increase the accuracy of prediction, at the lowest recorded height [69]. Moreover, to investigate the impact of RES intermittency, random Gaussian noise is added to the baseline deterministic profiles. Respective stochastic parameters for wind and PV during MS1/2 are found from their historical hourly generation potential [70].

## 2.4 Power Flow Calculations
### 2.4.1 Nanogrid Operation

The LFS is implemented for operational control simulation of the design alternatives evaluated. For any hourly timestep $t$, the net electricity demand load $P_{NEL}{}^{n,t}$ at node $n$ is calculated from the hourly electricity demand $ED^{n,t}$ and on-site renewables generation $P_{res}^{n,t}$ (accounting for inverter and converter efficiencies). EH loads are prioritised over BESS charging, limited by installed capacity $\theta_{eh}^{n,t}$, hourly maximal output $P_{eh,max}{}^{n,t}$, and power available from total generation $P_{eh,res}{}^{n,t}$. The remaining power $P_{res,PL}^{n,t}$ and heating demand $P_{NHL}^{n,t}$, if any, dictate BESS operation at each node, which is prioritised over PP distribution to minimise distribution losses. The energy flows into the BESS capacity $\theta_{ba}^{t}$ are then determined from their technical parameters, namely charging $C_{in}$ and discharging $C_{out}$ rates as well as maximum nominal input $P_{ba,in,max}^{n,t}$ and output $P_{ba,out\,max}^{n,t}$ power. They are also bounded by the minimum $E_{ba,min}^{t}$ and maximum $E_{ba,max}^{t}$ storage capacities for the selected BESS. The battery storage level $E_{ba}^{t}$ is thus computed as the sum of the stored energy from the previous timestep $E_{ba}^{t-1}$, storage energy flow $P_{ba}^{t}$ and any potential additional installed storage capacity during that period $E_{ba,new}^{t}$. Note that newly purchased batteries are assumed to arrive *fully charged,* and conversion losses are accounted for via the battery roundtrip efficiency parameter $\eta_{ba}$. Remaining excess energy $P_{eg}^{n,t}$ after nanogrid operation is then used to engage in PP operation

### 2.4.2 PP Micro-grid Distribution

Distribution requirements are calculated at the entire system level from each node, which are classified as either $N_{deficit}$ or $N_{surplus}$, based on the obtained energy balance. The system wide unmet load $P_{ul,es}^{t}$ and available extra generation $P_{eg,es}^{t}$ are thus used to roughly estimate micro-grid distribution requirements $P_{dist,es}^{t}$ at an hourly resolution. The micro-grid power distribution potential is limited by the capacity of the cables connecting the different units $\theta_{PP\,mg}^{t}$, and is capped at no more than 30% of the system wide electricity load $P_{dist,es,max}^{t}$, aiming to balance network reliability with increasing infrastructure cost, as shown in (2).

$$P_{dist,es}^{t} = \min(P_{ul.es}^{t}, P_{eg,es}^{t}, \theta_{PPmg}^{t}, P_{dist,es,max}^{t}) \qquad (2)$$

Furthermore, to tractably approximate performance, it is assumed that the distribution of power is spread *homogenously* across the micro-grid network, as both loads, and generation should be well distributed spatially in light of migration compatibility objectives. More specifically, representative power ($P_{dist}^{t,n} = \frac{P_{dist,es}^{t}}{N}$) and length ($L_{dist}^{t,n} = \frac{L_{cb}^{t}}{N}$) requirements for PP distribution per each node $n$ are developed for preliminary loss calculations. Dwellings are assumed to be connected via 3-phase transmission rated at 220/380 V through low-cost copper cables [17]. Interconnection resistance per node $R_{dist}^{t,n}$ is thus found from the manufacturer reported reference value and tuned to the actual operating conditions through the material temperature coefficient $\alpha_{cb}$. Cable resistance is used to estimate energy system losses, $P_{dl,es}^{t}$ according to (3), neglecting reactive power and only focusing on thermal losses associated with the induced current $I_{dist}^{t,n}$ in each line. Finally, distribution loss proportion $DL\%$ is found from (4), seeking to capture overall network efficiency. The balance of power flows in the network after PP operation is ensured by computing the final excess electricity generation $P_{res,exc,pp}^{t}$, unmet electricity load $P_{uel,es,ppp}^{t}$ and/or unmet heating load $P_{uhl,es,ppp}^{t}$. These values then dictate utility grid interactions, as detailed in the following section.

$$P_{dl,es}^{t} = \sum_{n \in N} 3R_{dist}^{t,n} I_{dist}^{t,n}{}^{2} \qquad (3)$$

$$DL\ (\%) = 100\frac{P^t_{dl,es}}{P^t_{dist,es}} \tag{4}$$

### 2.4.3 Utility Grid Interaction and Coal Requirements

The potential to buy electricity from the distribution grid during MS2 is considered in this section. Interconnection is simulated at a system level through simple heuristics (i.e. market engagement is *not explicitly* considered) to focus on flexibility strategies. Interactions with the main distribution grid, limited by line capacity $\theta^t_{Grid}$, unmet heating load $P^t_{uhl,es,ppp}$, and availability $\rho_{grid}$, are calculated based on system energy balance, as shown by (5) below. Note that $\rho_{grid}$ is a binary variable representing grid availability for that timestep, sampled from a *beta* distribution. Case (i) then represents timesteps when blackouts are happening so no interactions happen. Case (ii) shows situations when demand has not been fully met, thus electricity is purchased from the grid. For case (iii), on the other hand, the excess generation after meeting micro-grid loads $P^t_{res,exc,pp}$ is sold back to the main grid. Note that excess generation $P^t_{ex}$ which cannot be captured by grid interactions is assumed wasted at this stage. Please refer to 9.1 for the complete power flow model equations.

$$P^t_{grid} = \begin{cases} 0 & \rho_{grid} = 0\ (i) \\ \min(\theta^t_{Grid}, P^t_{eh,max}, P^t_{uhl,es,ppp}) & P^t_{uhl,es,ppp} \geq 0, \rho_{grid} = 1\ (ii) \\ -\min(\theta^t_{Grid}, P^t_{res,exc,pp}) & P^t_{uhl,es,ppp} \leq 0\ ; \rho_{grid} = 1\ (iii) \end{cases} \tag{5}$$

## 2.5 Emissions Generation

Based on the above, the heating demand not met via EH, and therefore mass of coal required for the *entire nomadic community*, is calculated as shown in (6)-(7). $P^t_{coal,es}$ captures the thermal energy requirements after PP distribution $P^t_{uhl,es,ppp}$ and grid interactions $P^t_{grid}$. The associated coal mass $M_{coal}{}^t$ is estimated from its average energy density $\mu_{local\ coal}$ and thermal efficiency of the traditional stove used for combustion $\eta_{coal\ stove}$. Energy system carbon footprint over time is estimated from the mass of raw coal burned and grid purchased electricity according to (8), using their respective weighted average emissions factors ($EF_{grid}$ and $EF_{coal}$).

$$P^t_{coal,es} = \max\ (P^t_{uhl,es,ppp} - P^t_{grid}, 0) \tag{6}$$

$$M_{coal}{}^t = \frac{P^t_{coal,es}}{\eta_{coal\ stove}\ \mu_{local\ coal}} \tag{7}$$

$$CO_2{}^t_{eq.emissions,es} = EF_{grid} P^t_{grid} + EF_{coal} M_{coal}{}^t \tag{8}$$

## 2.6 System Costs

The model presented thus far ultimately seeks to calculate system costs at a monthly $m$ resolution $C^m_{es}$ in each uncertainty scenario. As shown in (9) below, the formulation accounts for investment costs $C^m_{inv}$, operational costs $C^m_{opex}$, any expenses (or revenues) incurred through interactions with the distribution grid $C^m_{grid}$, as well any potential penalty (or credit) associated with generated emissions $C^m_{CO2}$. Representing liquidity and economic limitations faced by these communities, a constraint is formulated to cap monthly household energy system expenditure at the cluster level, as shown by (10). The modelling of each of these cost components is discussed in more details below.

$$C^m_{es} = C^m_{inv} + C^m_{opex} + C^m_{grid} + C^m_{CO2} \tag{9}$$

$$C^m_{es,max} \leq C^m_{es,max} N_n \tag{10}$$

### 2.6.1 Investment Costs

Investment costs in each timestep, as shown in (11), are calculated from the communication $CC_{n,i}{}^m$, infrastructure $IC_{es}^m$, expansion $EC_{n,i}$, and replacement costs $RC_{n,i}$, net the residual value (if any) $SV_{n,i}$ recorded across all nodes and technologies evaluated. Communication infrastructure costs $CC_{es}^m$ required for the safe operation and control of the PP microgrid are preliminarily estimated through a simplified model based on system distribution requirements $P_{dist,system}^m$ and a cost constant factor $c_{cc}$, as shown in (12). Assuming a 2G/3G/4G base station is already present, the communication infrastructure needed is thus simply an add-on support enabling peer-to-peer communication among nanogrid units. They are considered as a one-off investment incurred only in timesteps where $P_{dist,es}^m$ is higher than monthly distributed electricity in all previous periods $P_{dist,es}^{m,max,past}$, representing the need for infrastructure upgrade - neglecting the effect of EoS. Similarly, $IC_{es}^m$ – which represent the need for additional cabling to connect the various nanogrid units- are recorded for timesteps when the current cluster radius is higher than that recorded in any of the previous simulation months $r_{cluster}{}^{max,past}$. It is thus assumed that cabling infrastructure purchased at any point during project life will remain available for connection if needed again. The cost of required extra cabling is computed according to (13), using a unit length cost factor $c_{ic}$ obtained from regional manufacturers [71], invariant with time or scale.

The formulation of $EC_{n,i}$ in (14), on the other hand, accounts for both of economies of scale (EoS) and cost of transportation in the targeted settings. The estimations for the investment constant $K_i$ and EoS factors $\alpha_i$ are, therefore, slightly reduced and increased respectively in MS2 as compared to the MS1 [72], and based on specifications from regional manufacturers per component [73-79]. Coal stove capacity is an exception to this, where it is assumed that no starting stage investment is required, although replacement costs are incurred every 7-years [80]. As such, stove capacity is not explicitly optimised, and it is simply assumed one will be needed per family at a fixed (and nearly negligible) cost over time. Any other given capacity decision $\theta_{n,i}^{m,new}$ is expected to yield additional power electronics requirements $\theta_{n,pe}^{m,new}$, assumed to be proportional to expansion magnitude. This is based on reported ratios of nominal power electronic to system component $i$ capacity, $\varepsilon_{pe,i}$, averaged over several micro-grids from the literature [14, 19-21, 81, 82]. A safety factor $sf_{pe}$ of 1.3 is included to further ensure no undersized components, as shown in (15), yielding total power electronics costs $PEC_{n,i}{}^m$ through a constant cost factor $c_{pe,i}$.

Replacement costs $RC_{n,i}$ are incurred when a system component $i$ reaches the end of its predicted operational life $L_i$, using the same parameters as (14). As such, $RC_{n,i}$ may be experienced for multiple technologies in the same month- in contrast to $EC_{n,i}$, which is limited to a single system component $i$ per timestep. The asset salvage value $SV_{n,i}$ is included for cases where capacity abandonment decisions are made, or for NPC calculations at the end of each simulated episode. As shown in (16), it is a function of remaining estimated functional hours $L_i^\theta$ for a component with an original maximal lifetime $L_i{}^{max}$, and initial investment cost based on the salvaged capacity amount $\theta_{sv,n,i}^m$, corrected through a depreciation factor $k_i^{salv}$ which is smaller than 1.

$$C_{inv}^m = CC_{es}^m + IC_{es}^m + \sum_{i \in I}\sum_{n \in N} EC_{n,i}{}^m + RC_{n,i}{}^m - SV_{n,i}{}^m \tag{11}$$

$$CC_{es}^m = \begin{cases} c_{cc}(P_{dist,es}^m - P_{dist,es}^{m,max,past}) & P_{dist,es}^m > P_{dist,es}^{m,max,past} \\ 0 & P_{dist,es}^m \leq P_{dist,es}^{m,max,past} \end{cases} \tag{12}$$

$$IC_{es}^m = \begin{cases} c_{ic}(r_{cluster}{}^m - r_{cluster}{}^{max,past}) & r_{cluster}{}^m > r_{cluster}{}^{max,past} \\ 0 & r_{cluster}{}^m \leq r_{cluster}{}^{max,past} \end{cases} \tag{13}$$

$$EC_{n,i}{}^m = K_i(\theta_{n,i}^m - \theta_{n,i}^{m-1})^{\alpha_i} + PEC_{n,i}{}^m = K_i(\theta_{n,i}^{m,new})^{\alpha_i} + PEC_{n,i}{}^m \tag{14}$$

$$PEC_{n,i}{}^m = c_{pe,i}\theta_{n,pe}^{m,new} = c_{pe,i}\,sf_{pe}\,\varepsilon_{pe,i}\,\theta_{n,i}^{m,new} \tag{15}$$

$$SV_{n,i}{}^m = k_i^{salv}\frac{L_i^\theta}{L_i^{max}}K_i(\theta_{sv,n,i}^m)^{\alpha_i} \tag{16}$$

### 2.6.2 Operational Costs

The operational costs $C_{opex}^m$ detailed in (17)-(21) below account for expenses due to heating via a traditional coal stove $C_{coal}^m$, maintenance of fees associated with each energy system component $C_{om}^m$ and the economic value of unmet electricity load $C_{ul}^m$. Expenses associated with space heating are found through the required monthly coal mass $M_{coal}{}^m$ and the local average price by weight $c_{coal}$ – assumed invariant with time [80]. O&M costs are found as a function of total energy system installed capacity $\theta_{i,es}{}^m$ and respective periodic estimated maintenance fee $c_{om,i}$ for each component $i$, using recently reported average values for small scale systems [83]. To compute the economic value of energy not served, the total monthly unmet load $ED_{ul,es,final}^m$ for the entire system is multiplied by the curtailment fee $c_{ul}$, as shown in (20). It is assumed that a *5 % monthly shortage is allowed*, thus no charges are associated below that level. The value for $c_{ul}$ is estimated through the Willingness to Pay (WTP) approach, which relates perceived marginal benefits to the maximal price a customer would be willing to pay for a service [84]. It is, therefore, assumed equivalent to the local cost of diesel which would be needed to supply the unmet load – neglecting generator investment cost and assuming standard conversion efficiencies. Cash flows resulting from grid interactions can be either positive or negative, based on the monthly energy balance $P_{grid}^m$, and are computed using either the reported discounted tariff $c_{grid}$ or subsidized price received $p_{grid}$.

$$C_{opex}^m = C_{coal}^m + C_{om}^m + C_{ul}^m \tag{17}$$

$$C_{coal}^m = c_{coal}\,M_{coal}{}^m \tag{18}$$

$$C_{om}^m = \sum_{i\in I} c_{om,i}\theta_{i,es}{}^m \tag{19}$$

$$C_{ul}^m = c_{ul}\,ED_{ul,es,final}^m \tag{20}$$

$$C_{grid}^m = \begin{cases} c_{grid}P_{grid}^m & P_{grid}^m \geq 0 \\ p_{grid}P_{grid}^m & P_{grid}^m < 0 \end{cases} \tag{21}$$

### 2.7 System Mass

The total system mass $M_{es}^m$ can be found from (22) for any month $m$ as the sum of the weight of the installed renewables $M_{res}^m$, batteries $M_{batt}^m$, power electronics $M_{pe}^m$, and cabling length required for connection $M_{cb}^m$. To reflect mobility considerations associated with a migratory lifestyle, an upper limit on energy system weight $M_{es,max}^m$ may be imposed, captured by (23).

$$M_{es}^m = M_{res}^m + M_{batt}^m + M_{pe}^m + M_{cb}^m \tag{22}$$

$$M_{es,max}^m \leq M_{es,max}^m N_n \tag{23}$$

## 2.8 Uncertainty Simulation

Geometric Brownian Motion (GBM) models are developed to estimate stochastic evolutions of the most important sources of uncertainty integrated in the energy system planning problem. This form of stochastic model is chosen as it can represent, for any variable $x$, both overall long-term trends (if any), and the potential for tail event or low probability scenarios. This is deemed highly compatible with most of the uncertainties considered in this work. It should be noted, however, that if available data were to suggest otherwise, alternative, and more complex stochastic model formulations should be explored outside this proof-of-concept work. Mathematically, the model involves drift $\mu_x$ and volatility $\sigma_x$ variables, respectively. $dW_t$ models the random Wiener process, or stochastic deviations from the mean as sampled from a normal distribution, and $dt$ the time resolution implemented. These values, typically based on historical data, allow one to fine-tune the stochasticity of the model, while not limiting the potential total evolution scenarios [85-87]. They can then be used to simulate uncertainty for any variable $x$ as shown by (24-25) below.

$$dX^t = \mu_x X^t \, dt + \sigma_x X^t dW_t \tag{24}$$

$$X^t = X^0 \exp((\mu_x - {\sigma_x}^2/2)t + \sigma_x W_t) \tag{25}$$

## 2.9 Energy System Model Overview

Figure 3 captures the interaction among different model components detailed in the former sections, leading to calculations of total monthly net system cash flow ($NCF^t = C_{es}^m$) and emissions $CO_{2\,eq.emissions}^m$ across different simulations. Changes or upgrades to the energy system state are determined by the DRL agent policy captured by the green box, as detailed in 3.1.2, impacting both investment and operational costs incurred in the remaining simulation steps. For the baseline rigid systems, on the other hand, cost calculations are based purely on uncertainty realizations.

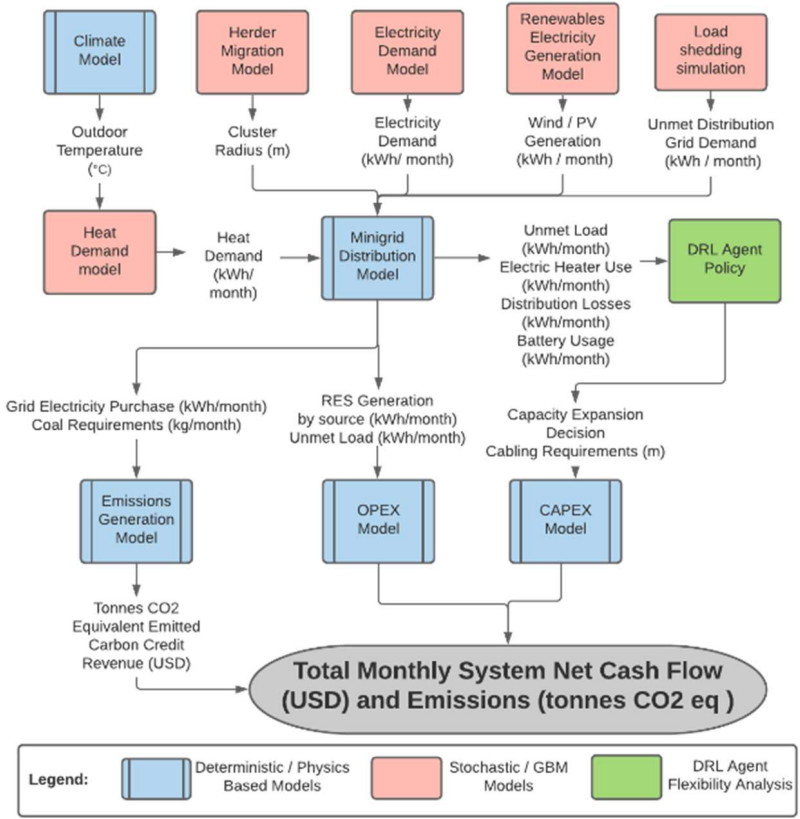

*Figure 3: Model flowchart showing complex interactions among system components and simulations*

# 3 Methodology
## 3.1 Rigid Stochastic Capacity Planning Model
### 3.1.1 Mathematical Formulation

This section introduces the generic formulation of the rigid capacity planning model under uncertainty. It is later used to benchmark the performance of the proposed DRL approach on an example mobile energy supply system. Considering a discrete finite time horizon planning period of $T$, with discount rate $\lambda$, let $\xi = (\xi^1, \xi^2, \ldots \xi^T)$ be a scenario of uncertainty, where $\xi^t$ is a vector capturing the uncertainty observed in period $t$, thus able to account for multiple uncertainty sources and relevant variables. $S$ defines the set of all possible uncertainty scenarios $s$ used for evaluation, assumed to be finite at $S_s$ for simplicity, with equal probability of occurrence. The baseline single objective rigid stochastic planning model for the mobile energy system considered can thus be formulated as shown by (26)-(28). $\theta_i^t$ represents the installed nominal capacity for various system components $i$, invariant with time $t$. $\mathcal{R}^t(\theta_i^t, \xi_s^t)$ is the revenue function for the installed technologies and $\mathcal{C}^t(\theta_i^t, \xi_s^t)$ the cost function for the same technologies in period $t$. Initial investment costs for each of these components $CAPEX_i$ are included as a zero-stage decision, while periodic energy system costs under uncertainty $C_{s,es}^t$, detailed in 2.6, are captured by (27).

$$\min ENPC = \frac{1}{S_s}\sum_{s=1}^{S}(CAPEX_i + \sum_{t=1}^{T}(\frac{1}{1+\lambda})^t[\mathcal{R}^t(\theta_i^t, \xi_s^t) - \mathcal{C}^t(\theta_i^t, \xi_s^t)])) = \frac{1}{S_s}\sum_{s=1}^{S} NPC_s \qquad (26)$$

$$C_{s,es}^t = \mathcal{R}^t(\theta_i^t, \xi_s^t) - \mathcal{C}^t(\theta_i^t, \xi_s^t) \qquad (27)$$

$$\forall\, t \in T, s \in S, i \in I \tag{28}$$

### 3.1.2 Optimisation

The optimal sizing of the baseline inflexible systems is found through the genetic algorithm (GA) implementation in Python [88]. GA is a well-known and widely used approach to global optimisation compatible with mixed continuous/discrete variables, inspired from the selection process found in natural biological systems [89, 90]. Solutions are iteratively improved within the design space through operations such as crossover, mutation, and fitness selection to try and imitate the natural evolution process. Initially, candidate solutions for different baseline mobile energy systems (or "chromosomes") are generated probabilistically with a random component across the design space. This helps explore a wide range of capacities for the different technological option baselines considered, as clarified in 5.1. These candidates are then refined and or/discarded sequentially through the genetic operations mentioned. The fitness function this process seeks to optimise is given by (26), or the minimisation of ENPC over 2,000 (fixed) simulated scenarios. A penalty function is implemented to ensure constraint-feasible solutions, as presented in [91].

## 3.2 DRL Flexible Capacity Planning Model

### 3.2.1 Background and Notation

DRL is designed to make decisions in Markov Decision Process (MDP) and is intuitive in nature. It involves an agent interacting with a system (or environment) over time through a large number of iterations as part of a sequential decision-making problem, with the aim of reward maximization [92]. Generally, at each time step $t$, the agent finds itself at state $s^t$, and selects an action $a^t$ following a decision-making policy $\pi$. As a result of taking the action, the agent transitions to a successive state $s^{t+1}$ receiving a corresponding scalar reward $R^t$, which are governed by the environment's dynamics and reward function $\mathcal{R}(s^t, a^t)$, respectively [93]. This repeated interaction, has the objective to improve the accuracy of expected value estimations for either each state $V_\pi(s^t)$ or state-action pair $Q_\pi(s^t, a^t)$. They refer to the total expected value of being in a particular state or the *decomposed* expected value by each action available to the agent in that state, as shown in (29) and (30) respectively. Based on the above, the agent can map and rank state-action pairs through their relative value. Ultimately, the aim is to develop an optimal decision-making policy $\pi^*$, as captured in (31), which maximizes *discounted* (through γ) accumulated rewards per episode. In a flexibility planning application, an episode is equivalent to a simulation of the full project horizon.

$$V_\pi(s^t) = \mathbb{E}\left[Q_\pi(s^t, a^t)\,\middle|\, s^t\right] \tag{29}$$

$$Q_\pi(s^t, a^t) = \mathbb{E}[R_\pi(s^t, a^t)] \tag{30}$$

$$\pi^* = \arg\max_\pi \mathbb{E}_{\tau \sim \pi(\tau)}\left[\sum_t \mathcal{R}(s^t, a^t)\right] = \arg\max_\pi \mathbb{E}_{\tau \sim \pi(\tau)} \sum_t \gamma\, R^t \tag{31}$$

The optimal policy (or flexibility strategy) is then typically iteratively approximated inside a DNN through either value based or policy-based methods. The DNN often consists of several layers to improve the granularity of value (or policy) function representation over the training process (as shown inside the agent in Figure 4), with optimal architectures varying based on the problem at hand [94]. Approaches belonging to the former, such as Q-learning, normally follow the trajectory of the highest value estimates to formulate the optimal policy. This allows for off-policy learning, as each update can use data collected during all training interactions, although potentially leading to estimation bias and improper credit assignment [55, 95]. Policy based methods, on the other hand, parametrize the policy directly by creating an underlying action selection probability distribution, making them generally more effective in high-dimensional stochastic environments [96]. Most modern DRL algorithms, nonetheless, implement hybrid actor-critic formulations to combine aspects of both for state-of-the-art results on benchmark tasks [53, 93, 97, 98], motivating their selection for this work. Generally, they are defined by an actor which determines which action to take under a policy, and a critic which approximates how good the actor's selection was based on value function estimates. The actor-critic architecture implemented in this study is further detailed in 3.2.4.

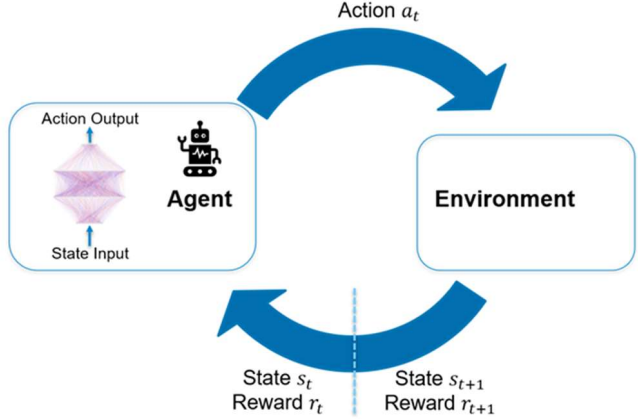

*Figure 4: Graphical overview of DRL agent-environment interaction*

### 3.2.2   DRL Problem Formulation

A DRL based approach is used to evaluate and optimize an implicit form of *decision rules* to act as a triggering signal for making a particular flexibility decision during system operation, as captured by a policy $\pi$ within a DNN. The major difference compared to alternative methods presented in 1.2.3 is that $\pi$ can be characterized by a much larger set of parameters, and thus actions need not be restricted to a static pre-determined form. The primary steps required to formulate an energy system design and planning optimisation through the DRL Flexibility Analysis framework are problem formulation, uncertainty recognition, and action space creation – further discussed in 0.

More generally, it is assumed that the condition of the energy system for any $t \in T$ can be defined by a state observation, $s^t \in \mathbb{X} \subseteq \mathbb{R}^{n_x}$, where $s^t$ captures the installed capacities for each component $i$, in addition to other factors. Based on the information given to stakeholders at each evaluation time step, they can sequentially select a system evolution decision, $a^t \in \mathbb{A} \subseteq \mathbb{Z}^{n_a}$, changing the capacity of a component if deemed necessary. Considering the targeted objective, a reward function, $R: \mathbb{X} \times \mathbb{A} \times \mathbb{X} \to \mathbb{R}$, can be defined which captures the impact of different available actions over time. This is based on a series of periodic revenue $\mathcal{R}^t(s^t, \xi^t)$ and cost $\mathcal{C}^t(s^t, \xi^t)$ functions, approximated through environment interactions, as well as a *deterministic* expansion cost function $\mathcal{H}^t(a^t, s^t)$, including a zero-stage decision on starting system configuration.

The energy system state is then updated as a function of the previous state, action selected, and realisations of stochastic variables. As such, the system is identified to hold the Markov Property, and can be formulated within the MDP framework. The objective, as for the baseline design, is minimisation of ENPC (or maximization of negative reward) through the development of a system evolution policy, $\pi: \mathbb{X} \to \mathbb{A}$. Note that $a^t$ is then obtained from $\pi$ by using $s^t$ as an input, as shown in Figure 4. The generic model for DRL based energy system flexible capacity planning is defined according to (32), and can be implemented on the same scenarios $S_s$ as in (26) above. Furthermore, as $\pi(\emptyset, \cdot)$ is approximately captured by a DNN with parameters $\emptyset \in \mathbb{R}^{n_\emptyset}$, the capacity planning problem simplifies to obtaining optimal network parameters, or $\emptyset^*$.

$$\mathcal{P}(\pi) := \begin{cases} \max_\pi \mathbb{E}_\pi[-NPC] \\ s.t. \\ NPC_s = \sum_{t=0}^{T} \left(\frac{1}{1+\lambda}\right)^t R^t{}_s \\ R^t{}_s = C^t_{s,es} = \mathcal{R}^t(s^t, \xi_s^t) - \mathcal{C}^t(s^t, \xi_s^t) - \mathcal{H}^t(a^t, s^t) \\ s^{t+1} = f(s^t, a^t, \xi_s^t) \\ a^t = \pi(s^t) \\ \forall\, t \in T, s \in S \end{cases} \quad (32)$$

### 3.2.3 Environment Creation

Central to the applicability of DRL flexibility analysis is the development of an agent-environment interaction framework which is representative of the targeted objective. The Open AI gym format environment [99] is implemented to for the optimization of the mobile system given by (32). In this context, the system evolution strategy is primarily dependent on the periodic demand for different energy services, installed generating capacities and migratory patterns, among other factors. With the aim of reflecting this, the states for the agent are defined through a box observation set as:

$$s^t = \left[ED^t, HD^t, r^t_{cluster}, L_{cb}{}^t, \theta_{pv}{}^t, \theta_{wind}{}^t, \theta_{eh}{}^t, \theta_{bess}{}^t, ED^t_{ul,es,final}, M^t_{es}\right] \quad (33)$$

where $ED^t$ and $HD^t$ represent the electricity and heating stochastic demand realizations in timestep $t$, respectively. $r^t_{cluster}$ and $L_{cb}{}^t$ measure the uncertain spatial dispersion of the mobile system investigated and are used in cabling investment and distribution loss calculations. As in the former section, $\theta_i{}^t$ represents the installed *nominal* capacity for each system component $i$ considered. $ED^t_{ul,es,final}$ is included to stimulate the explicit consideration of unmet load, outside of charges incurred, while $M^t_{es}$ is used to account for increases in system mass as it evolves over time. Budget constraints are implicitly evaluated based on recorded rewards. All state variables are normalized based on the maximum and minimum values obtained for them over 10,000 simulated scenarios, independent of the environment. Furthermore, although not shown in (33), time limit awareness is included to improve end of episode policy performance and training stability [100].

The decision-making possibilities for the agent are also selected with the aim of reflecting the ones available to a ger community, through a multi-discrete action set ($\in A_{i,j}$). Through this formulation, the DRL decision space is such that, at any given time step, the agent may decide to leave the energy system configuration unchanged. On the other hand, based on estimated value function and policy, a decision enabled by flexibility may be made to alter the state of the microgrid. In that case, the multi-discrete set means a decision is made on which technology $i$ to act on, and whether to pursue capacity abandonment, expand by 500W or expand by the (rounded) maximum number of modules $\theta^t_{Budget}$ possible under the budgetary constraints $C^t_{Budget}$ for that period. As actions are selected at the entire system level, a simple heuristic is implemented to determine which node to implement the capacity adaptability decision on. For expansion decisions on solar PV, wind or BESS, additional component capacity is assumed to be installed at the node with the largest unmet load in the previous month $ED^{n,t-1}_{uel,max}$. Abandonment actions or EH expansions, on the contrary, are allocated to the node with the largest *excess* generation $P^{n,t-1}_{res,excess}$ prior to implementing the flexibility option.

Figure 5 below captures the internal evaluation process, going from state observation (left) to action selection (right) using a DNN parametrized agent policy, where components are evaluated within the environment technological boundaries. The iterative interaction feedback loop guiding policy updates during training, is visualized through the changes in system state, reward and policy driven by a capacity expansion decision made at time $t$ on $t+1$.

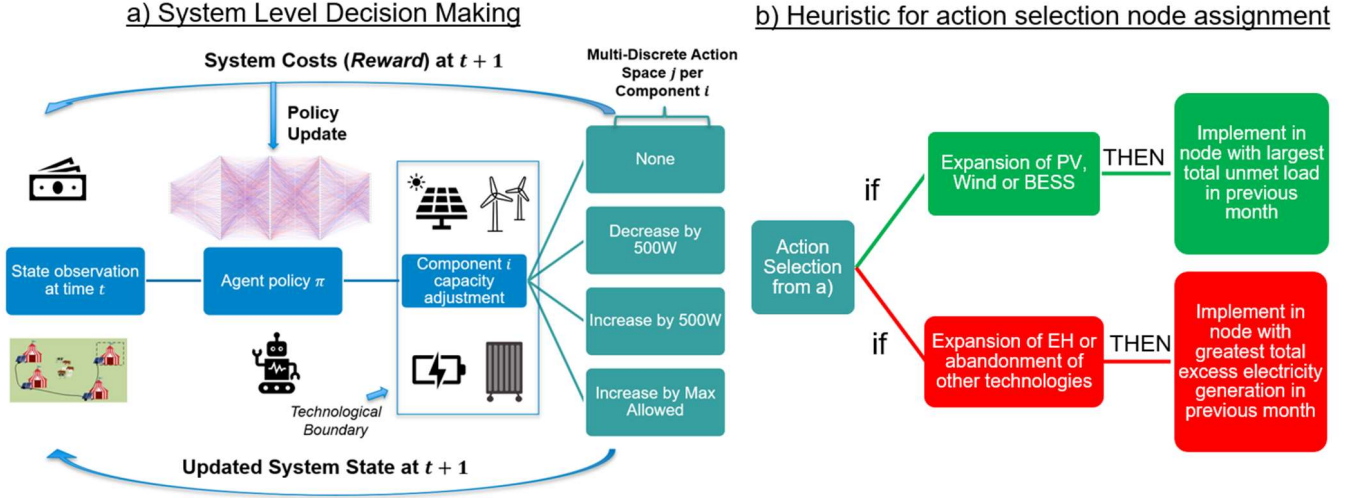

*Figure 5: a) Graphical overview of DRL system level decision making process going from microgrid state (left) to action selection (right). b) Simple heuristic used for allocation of system level DRL actions to individual microgrid nodes, either the ones with either largest unmet load in previous month (top) or greatest total excess electricity generation (bottom).*

### 3.2.4 Algorithmic Approach

Considering the dimensionality and computational expense of the problem tackled, the ACER (*Sample Efficient Actor Critic with Prioritized Experience Replay*) algorithm is implemented [101]. Each component is captured by a separate neural network, allowing different strategies to be implemented in policy as compared to value function updates. Experience replay allows agents to focus on the interactions most significant to experienced reward, improving convergence rate during training [102]. Sample efficiency was motivated by the complexity of the model, evaluated with different timestep resolutions for operation and decision making, meaning each environment episode simulation is relatively expensive computationally. This is achieved in ACER by combining several ideas from previous works with novel developments.

More specifically, multi-step impacts on state-action value function $Q_{ret}(s^t, a^t)$ approximations for each generated trajectory are found recursively using the Retrace algorithm [103], as shown in (*34*). Multi-step returns used allow for significant reductions in policy gradient bias and learning iterations needed for the Critic component of the algorithm. A duelling architecture is implemented, through which the vectorized value function estimates $Q_\emptyset(s^{t+1}, a^{t+1})$ as well as the policy $\pi_\emptyset(a^t | s^t)$, are outputs of the DNN with parameters $\emptyset$. As shown in (29), the estimate $V_\emptyset(s^{t+1})$ is then easily found as the expectation of $Q_\emptyset$ under $\pi_\emptyset$. Additionally, note that $\gamma$ is mathematically equivalent to the financial discount factor $\frac{1}{1+\lambda}$ shown in the problem formulated in (32). It is used to in DRL to account for the different value of immediate rewards compared to long term ones. The truncated importance weight $\bar{\rho}^t$ under current behaviour policy $\mu$, is thus computed as shown in (*35*) [104].

$$Q_{ret}(s^t, a^t) = R^t + \gamma \bar{\rho}^{t+1}[Q_{ret}(s^{t+1}, a^{t+1}) - Q_\emptyset(s^{t+1}, a^{t+1})] + \gamma V_\emptyset(s^{t+1}) \tag{34}$$

$$\bar{\rho}^t = \frac{\pi_\emptyset(a^t | s^t)}{\mu(a^t | s^t)} \tag{35}$$

Policy updates are then conducted in a very similar manner to Trust-Region Policy Optimization [105], although using the running average policy network as a baseline instead. The policy is thus decomposed into a probability distribution $f$, and network capturing the properties of the policy distribution $\tau_\emptyset$, with the update split into two stages. In the first stage, a linearized Kullback–Leibler (KL) divergence constraint is combined with a standard loss minimization optimization. In the second stage, backpropagation is used to compute the derivatives with respect to the policy parameters, adopting a trust region based on the distribution $f$ and advantage function. The gradient of ACER is thus defined as shown by (*36*)36 below, where $c$ is the correction term coefficient for cases with very high variance. Please refer to the original publication for further details and definitions [106].

$$\hat{g}_{ACER}{}^t = \bar{\rho}^{\,t} \nabla_{\tau_{\emptyset(s^t)}} \log f(a^t | \tau_\emptyset(s^t))\, [Q_{ret}(s^t, a^t) - V_\emptyset(s^t)]$$
$$+ \mathbb{E}_{a \sim \pi} \left( \left[ \frac{\bar{\rho}^{\,t}(a) - c}{\bar{\rho}^{\,t}(a)} \right] \nabla_{\tau_{\emptyset(s^t)}} \log f(a^t | \tau_\emptyset(s^t))[Q_\emptyset(s^t, a^t) - V_\emptyset(s^t)] \right) \quad (36)$$

### 3.2.5 DRL Agent Training and Testing

The DRL agent described in the former section is trained using a budget of 6 million monthly decision timesteps, equivalent to 500,000 simulated full 30-year episodes. As performed on a standard laptop Intel(R) Core(TM) i7-8665U CPU @ 1.90GHz, the complete training process takes approximately 7-8 hours. All training scenarios are stochastically generated to be *i.i.d* from each other, thus leading to an *infinite* amount of potential scenario evolutions and state-action pairs. This was motivated by the need to develop a more generalizable adaptability strategy under different realizations of uncertainty. No exactly equal trajectories are thus possible inside the environment, and training process is fully non-anticipative, with all operational decisions based purely on recorded observations at a particular timestep. Hyperparameters are found through Optuna [107], with final training conducted using 50 environment steps per policy update, parallelized through 4 processes, a correction term of 12.3 and a replay ratio of 5. The replay ratio and denotes the average relative replay learning per policy learning – which only kicks in after 20,000 replay starts in this case.

The DNN architectures implemented for both the actor and critic component of the algorithm are fully connected, with input layers matching the problem feature space described above. The output layer for the actor is composed of twelve neurons, representing the multi-discrete action selection process depicted in Figure 5 above, whereby four independent SoftMax distributional operators are then used to characterize decisions for each component. Additionally, six hidden layers with 64 neurons each are implemented to approximate the complex stochastic functions in each. Crucially, it should be noted that this approximation derived from the DNN captured by (32) does not always guarantee the optimality of the results obtained - especially when considering the non-convexity of the function evaluated. Nonetheless, several studies suggest strong convergence properties for DRL to low lying local minima when using gradient-based techniques, which employ a number of heuristics to reduce the likelihood of obtaining sub-optimal solutions [53, 108-114]. The Stochastic Gradient Descent Momentum method [115] is thus implemented for DNN training parameter updates, using the exponentially moving average of the trust-region adjusted values from (36).

Two versions of the environment are developed to separate agent training and evaluation. In the training environment, a negative reward penalty (-USD 1,000) is included for timesteps where either the budgetary or mobility constraints presented in 4.2 are violated. This allows the DRL agent to implicitly learn their form, looking for system solutions which are able to minimise the occurrence of constraint violation, and is the most commonly used method. While more advanced approaches are available to ensure constraints are robustly met, particularly in safety critical applications [116-118], they are typically associated with a high computational expense and are complicated to implement. As such, given in this setting the constraints represent socio-cultural considerations rather than safety issues, the simpler constraint violation penalty method is deemed justified.

This penalty is then removed in the testing environment, allowing evaluation of the true ENPC rather than the one obtained through the modified training reward function. Furthermore, to evaluate flexible system performance as compared to the baseline designs, all results are reported using the same 2,000 simulated *out of sample* episodes. This means that while trained on the same stochastic model *parameters*, no complete scenario used in testing would have been encountered during training. The different time step resolutions implemented are summarised by Figure 6 below.

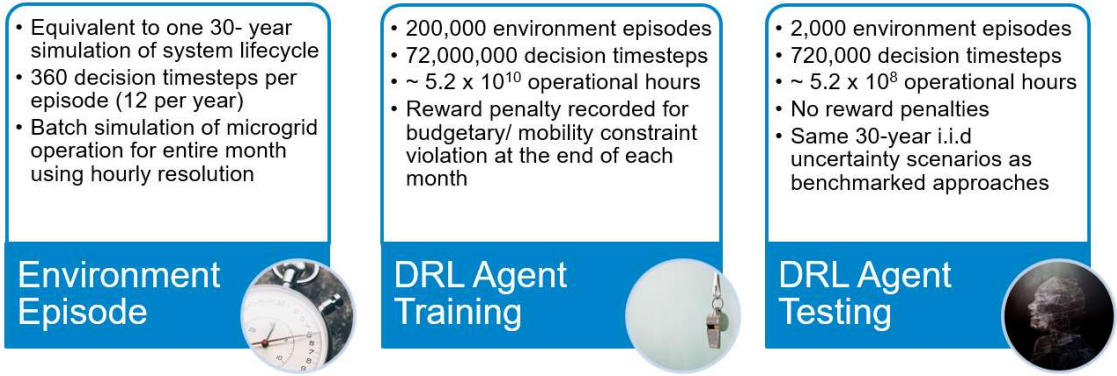

Figure 6; Different timestep resolutions implemented for the proposed DRL approach

## 3.3 Value of Flexibility and Plug and Play Operation

The total value of flexibility for the energy system considered $VoF^z{}_{Total}$, on any metric $z$, is captured by (37). In essence, it comes from the combination of tactical system planning flexibility $VoF^z{}_{Strategic}$, and the added value enabled by the PP control developments $VoF^z{}_{Operational}$. As shown in (38), it is thus estimated as the difference between the expected value for the flexible DRL PP solution $EV_{Flexible}{}^z$ and the best performing inflexible benchmark $EV_{BestBaseline}{}^z$. For $z = ENPC$, as formulated in 3.1 and 3.1.2, $VoF^{ENPC}{}_{Total}$ represents the maximum that should be paid to embed said flexibility in the energy system design. Furthermore, (39) is used to decompose the operational flexibility $VoPP^z$ generated by PP control possibilities from the DRL based decision-making flexibility proposed in this work. The value of operational flexibility is thus approximated from the difference in performance for baseline designs $EV^z_{No\ PP}$ as compared to ones integrating PP operation $EV^z_{With\ PP}$. While it is recognized the values for these flexibility sources may not be purely additive, they are considered separately in this work to determine whether or not the infrastructure investment required to enable PP operation is worthwhile in the targeted context. Decomposing relative performance improvements by source may also be beneficial for various policy-making objectives. It gives relevant decision makers important quantitative insights across different metrics $Z$. These can then be used to help develop market support instruments or incentive programmes, with the aim of maximizing added value from flexible PP operation for nomadic communities. Please note these estimates are calculated over 2000 *out of sample* scenarios, as further detailed in 5.1.

$$VoF^z{}_{Total} = VoF^z{}_{Strategic} + VoF^z{}_{Operational} \tag{37}$$

$$VoF^z{}_{Total} = EV_{Flexible}{}^z - EV_{BestBaseline}{}^z \tag{38}$$

$$VoF^z{}_{Operational} = VoPP^z = EV^z_{No\ PP} - EV^z_{With\ PP} \tag{39}$$

$$\forall z \in Z \tag{40}$$

# 4 Application Case Study
## 4.1 Background and Previous Work

Nomadic energy access and mobile energy systems are highly relevant in Mongolia, where it is estimated that 25-40 % of the national population continues to live a migratory lifestyle, normally moving between 4 and 30 times a year [119]. Changing climatic conditions, however, have increased the frequency of "zhuds" – extremely cold winters followed by extremely dry summers – making the traditional way of life more difficult for nomadic communities. This has resulted in an explosion of population migration to UB, where out of the approximately 1.4 million residents around 60% live in "ger areas" [119]. A "ger" refers to the traditional Mongolian herder dwelling, comprised of a basic wooden frame and a number of insulating layers to allow for mobility. On the other hand, "ger areas" represent the urban regions surrounding UB where the majority of these low-income communities now settle in the winter [120], and normally migrate to rural surroundings during the summer months, with annual movement ranges of around 7-8km [119]. Previous studies estimate nomadic community clusters are typically separated by 3.8km in the spring, 2.0km in the summer and 2.3km away in the autumn [121]. Figure 7 below graphically captures the migration patterns investigated, where MS1 represents operation in the area between UB and Hently in the summer, while MS2 is defined by migration to ger areas in UB during winter months, as in 2.1.

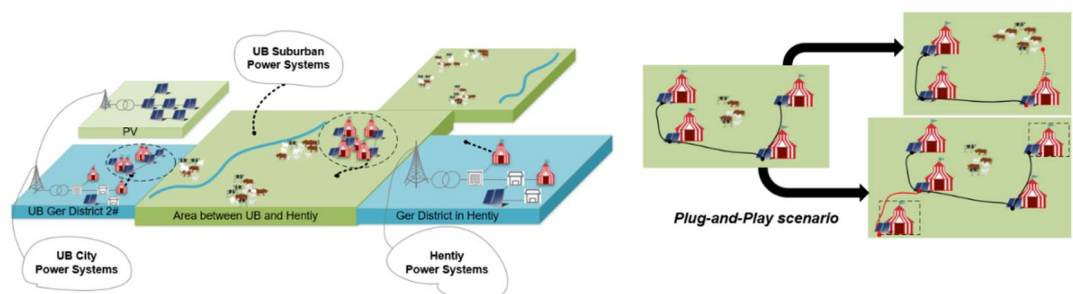

*Figure 7: Spatial boundaries for typical migration pattern (left) and example multi-ger PP configurations (right)*

### 4.1.1 Electrification

The national network infrastructure is largely unsuitable to provide reliable and sustainable electrification across the potential MS encountered. It currently relies primarily on coal (71%) and oil (25%) for centralised generation, with a large associated carbon footprint [122]. More specifically, during MS1 the Northern Energy System is the primary option for herder electricity connection [119]. Highly underdeveloped and unstable it mainly relies on imports from Russia [123]. It also only covers a fraction of the potential migration range, with 45% of rural gers unable to access electricity services consistently [124]. During MS2, on the other hand, gers are normally able to connect to the distribution grid in UB. This network, however, also experiences very frequent load shedding and blackouts, leading to limited energy security for ger communities [125].

In light of the above, household or community-based decentralised power supply systems are likely the most cost-effective option for achieving SDG7 in Mongolia [126]. In fact, more than 100,000 Small Home Solar (SHS) systems and 5,000 small wind systems (SWS) have already been distributed through the Renewable Energy and Rural Electricity Access Project (REAP), with capacities typically ranging from 20 to 100W per ger [123, 126]. REAP, however, was primarily based on capital investment subsidies and very low interest financing, thus not considering load evolution, mobility, or local resource assessment in capacity planning. Most of the systems currently installed are, as a result, either no longer in use or significantly undersized [126].

### 4.1.2 Heating Supply Decarbonisation

Electrification of nomadic communities has a very strong link with Mongolian government heating supply decarbonisation objectives, as residents of ger areas normally are unable to connect to the district heating network [80]. With the majority of the country in the extreme cold zone, there is substantial demand for space heating over a long winter season. As a result, it is estimated that ger areas contribute about 45% to 75% of UB's annual average fine particulate matter ($PM_{2.5}$) emissions [80, 127]. This is driven by the inefficient combustion of raw coal, biomass, or household waste, the dominant heating supply sources for low-income gers, making UB one of the worlds most polluted cities in winter months, with significant associated health issues [128]. Based on a median income of only around 350 USD $ger^{-1}$ $month^{-1}$, costs for space heating also contribute around 15-20% of annual ger expenditures [80, 129]. During winter, up to half of their limited monthly income may be spent on meeting heating demand alone, prioritising the cheapest source available.

Given the severity of and urgency of the air pollution issue, most previous works on ger energy systems seek to address decarbonisation of the heating supply under the extreme climatic conditions typically experienced [80, 120, 130]. Traditional coal-stoves are used in most gers, thus these studies normally focus on either benefits from using improved stove designs [131, 132] and/or better thermal insulation [80, 130]. Partly, this is because these works are centred on ger areas in the outskirts of UB where the grid capacity is a limiting factor to alternative electricity-based heating solutions, and the pollution is particularly severe. Most recently, however, [133] presented the experimental investigation and simulation of a solar PV generator combined with an EH for a single ger in UB, as compared to one using a traditional coal stove. The results from the study over a single year suggest a transition to electricity-based heating supply system can lead to improved indoor environmental quality, cost reductions, and significantly lower heat energy consumption.

## 4.2 Local Context Constraints

Given the mean monthly coal expenditure reported by [80], the total energy systems max expenditure $C_{es,max,ger}^{m}$ is limited to USD 110 $month^{-1}$ $ger^{-1}$. This results in approximately USD 20-60 $month^{-1}$ $ger^{-1}$ available for capacity expansion, depending on the respective coal expenditure, formulated at the entire 18-ger energy system level. Furthermore, it is well established that a typical 5-wall Mongolian ger weighs 250-400 kg, with an additional 75-100 kg for a standard coal stove and at least one 20kg bag of raw coal normally used during migration [120, 130]. Considering the above, an upper limit on total energy system weight $M_{es,max,ger}^{m}$ of 120 kg $ger^{-1}$ is preliminarily estimated.

## 4.3 Herder Migration

The assumptions used to model ger migration patterns and spatial dispersion uncertainty were developed and validated through direct engagement with a team of local consultants [72] and an interactive workshop with regional energy sector stakeholders [134]. Consequently, it was identified that 18 gers is representative of a typical ger nomadic community. Recent studies suggest that the average area occupied by such a community over the migration season is around 147 $km^2$, leading to an average seasonal cluster radius of **6.85 km** at project start [72, 121]**.** Furthermore, previous works suggest climate change is likely to lead to growth in overall migration distance and spatial dispersion over the 30-year horizon [119]. With the aim to capture how this uncertainty may influence optimal energy system configurations, mean annual growth rate and volatility are preliminarily estimated based on available data. The annual cluster radius $r_{cluster}^{y}$ evolutions are thus simulated according to (24)-(25). The resulting potentially large difference between projected (dashed line) and realised (solid lines) cluster spatial dispersion is shown in Figure 8, capturing the uncertainty space considered.

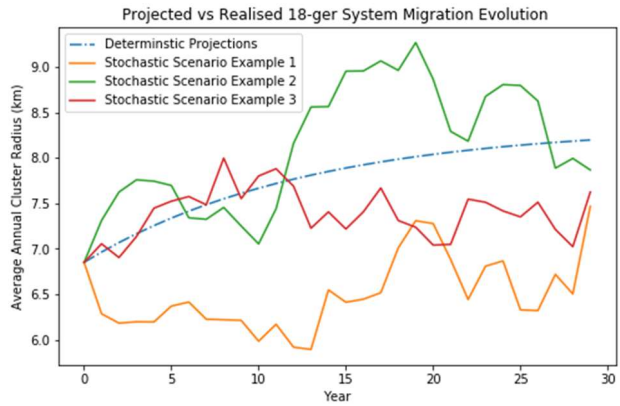

*Figure 8: Deterministic projections (**dashed line**) and 3 randomly selected example stochastic simulated scenarios (**solid lines**), showing the average annual (over the migration season) 18-ger cluster radius in km (y-axis), over the 30-year project horizon (x-axis). The projected migration growth is assumed to be driven by the climate change effects, as shown by the deterministic projections(**dashed line**), although it is recognized the magnitude of this impact is highly uncertain both in its long-term evolution and intra-year variability, as captured by the different stochastic example scenarios(**solid lines**).*

## 4.4 Electricity Demand (ED)

The baseline daily electricity load profile for a typical ger household was estimated through interactions with a local consulting team [72], given the limited literature data available. These values suggest most nomadic households continue to follow a more traditional way of life for now, typically only consuming enough electricity to operate a 20W lamp and 80W refrigerator. In winter months, however, the local climatic conditions as well as economic constraints mean that the refrigerator is not needed nor used. This assumption is maintained throughout project horizon, with starting electricity demand estimates of roughly 63 and 6 kWh month$^{-1}$ ger$^{-1}$ during the summer and winter months, respectively. Nonetheless, this load is likely to grow significantly over the next 30 years, with recent ones experiencing a 5-7% annual increase in electricity demand, driven by a combination of more productive uses of electricity integrated into nomadic lifestyles, higher indoor environmental standards, and the adoption of low-cost electric appliances [120, 123, 135].

Baseline load profile projections are thus formulated based on the integration of a 213W TV with a 30W antenna, more light bulbs, and a number of additional electric appliances (i.e., radios, cell-phone) by years 15 and 30, respectively [72]. The stochastic electricity demand model is defined from (24)-(25). The baseline daily load profiles for summer days are shown on the left in Figure 9 below at the starting point (Year 0), midpoint (Year 15) and end (Year 30) of the project. Long term evolutions and variation under uncertainty for different seasons, on the other hand, are captured on the right in Figure 9. For power flow calculations in 9.1, it is assumed that these same hourly load profile curves are maintained, the magnitude of which is scaled proportionally to the actual realized monthly demand as compared to the one projected at time of writing.

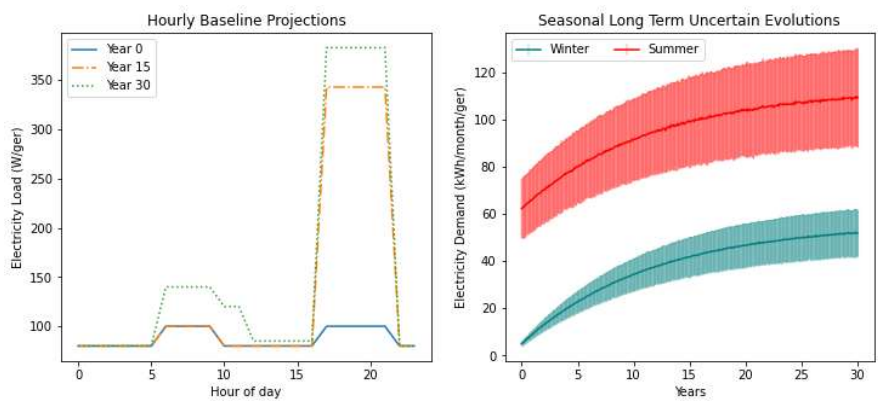

*Figure 9: Hourly projected load (left) and seasonal long term stochastic evolutions of electricity demand (right)*

## 4.5 Heating Demand (HD)

As a baseline for estimating heating demand (HD), it is assumed that all 18 gers in the systems evaluated have 5 walls with standard insulation. While the benefits of improved insulation for gers have been discussed in several other studies [129, 130, 133], the standard insulation case defined in these works is still representative of the majority of ger dwellings. Seasonal demand calculations are obtained through consultation with a local team [72] and are aligned with values found in the literature [80, 129, 130, 133]. Respective demand-temperature correlations are then used to extrapolate monthly profiles based on the assumptions given in 9.2, resulting in starting demand of 11,500 kWh/ year per ger. Volatility for heating demand is assumed to be approximately equivalent to that on year-to-year climatic conditions, found via standard regression analysis of the average *winter* temperatures between 1901-2020 [136]. The stochastic HD model is thus formulated yearly according to (24)-(25), including potential climate change induced growth [137-139]. The impact of this evolution and differences in potential Year 30 HD profiles is shown for 3 stochastic scenarios in Figure 10. As for ED above, the seasonal profile *distribution* is maintained across years simulated, and the value in each timestep scaled proportionally to the one at project start [72].

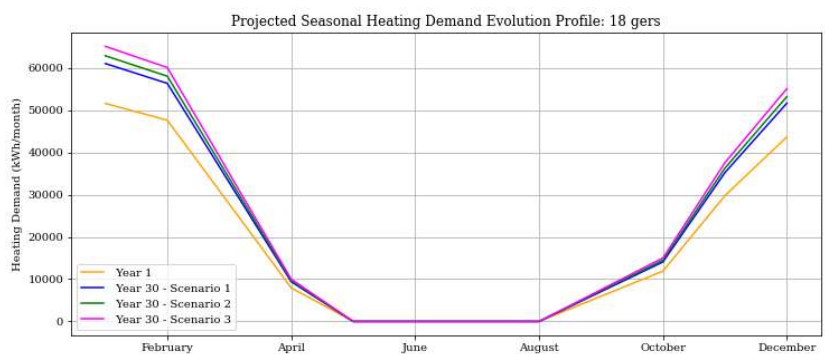

*Figure 10: Projected starting and end of project life heating demand monthly profiles for 3 scenarios*

## 4.6 Renewable Energy Generation

For the system sizes and representative locations considered in this manuscript, a mean annual capacity factor (CF) of 19.8% for solar PV is obtained, similar as reported in [133]. The initial estimate for annual wind capacity factor is slightly higher at 22.2%. This is lower than values from other wind generation sites in Mongolia, as the small-scale systems selected are operating at reduced wind speeds and conversion efficiencies compared to utility level ones [126, 135]. Still, both CF estimates are consistent with those which could be extrapolated from alternative data sources [140, 141] or with previous studies [122, 135, 142].

Please refer to 9.2 for the complete list, values and ranges of the parameters used in this study.

## 5 Results

### 5.1 Summary

The energy system design alternatives discussed in the remainder of this section are summarised in Table 1 below, aiming to capture the benefits of flexibility and a systems approach to planning for ger communities. The stochastically optimal rigid baselines results suggest the deployment of a SHS system for BD1 ($\theta_{pv}^{BD1}$ = 525 W/ger). This is approximately 5-10 times with what is currently available to herders, or roughly twice that obtained using commercial software such as HOMERPRO, which neglects load growth. Interestingly, the optimal PV capacity in BD2 is equal to that found in BD1, although with a roughly matching EH capacity (625 W/ger) for thermal energy supply. In BD2, therefore, the EH units are run almost exclusively with power purchased from the UB grid. BD3 and BD4 present very similar initial design capacities to their counterparts which are optimised for no PP connection (BD1 and BD2), albeit with reductions in BESS requirements of roughly 8% and 10%, respectively. System evolution for FD, on the other hand, varies based on scenarios, although with final capacities generally much greater (and diversified) than those obtained via rigid baselines.

*Table 1: Summary of ger energy system design alternatives evaluated in this study*

| Ger Energy System Alternative | Description |
|---|---|
| **Flexible Design (FD)** | 18-ger multi-energy PP *microgrid* design and adaptability strategy |
| **Flexible Nanogrid Design (FND)** | DRL based single nanogrid unit design and adaptability strategy |
| **Baseline Design 1 (BD1)** | Single nanogrid unit stochastically optimal rigid design with no EH |
| **Baseline Design 2 (BD2)** | Single nanogrid unit stochastically optimal rigid design with EH option |
| **Baseline Design 3 (BD3)** | 18-ger microgrid stochastically optimal rigid design with no EH |
| **Baseline Design 4 (BD4)** | 18-ger microgrid stochastically optimal rigid design with EH option |

The combined relative *out of sample* performance of the 18-ger energy system design alternatives optimised using the model from Section 2 are summarised by Table 2 below. They show that the FD yields improvements across all metrics estimating economic feasibility, sustainability, and energy security of alternatives, highlighting the importance of embedding capacity expansion flexibility under uncertainty. Note that, as presented in Section 0, $VoF$ refers to improvements achieved by FD for each metric as compared to the *best performing benchmark design on the same metric*. Furthermore, to capture decision makers with differing risk-profiles, Value at Risk (VaR) and Value at Gain (VaG) are evaluated for each solution, tailored to risk-averse and risk-seeking operators, respectively. These results suggest that FD may be highly valuable for ger communities' electrification along different risk and utility metrics, as discussed in more detail below.

*Table 2: Design decision making table for alternatives evaluated in this study over 2000 out of sample scenarios*

| *Metric* | *Objective* | BD1 | BD2 | BD3 | BD4 | FND | FD | VoF |
|---|---|---|---|---|---|---|---|---|
| **ENPC (USD)** | *Economic Feasibility* | 439,212 | 413,771 | 428,088 | 386,459 | 373,503 | 328,178 | **85,593** |
| **VaR, 5% (USD)** | *Economic Risk* | 492,314 | 516,332 | 480,058 | 473,062 | 419,512 | 367,421 | **124,893** |
| **VaG, 95% (USD)** | *Economic Upside* | 381,456 | 370,591 | 365,221 | 352,669 | 338,229 | 293,404 | **77,187** |
| **Lifetime Emissions (t. $CO_2$ Eq.)** | *Sustainability* | 10,133 | 14,169 | 9,808 | 13,375 | 4,937 | 4,601 | **5,532** |
| **Expected TUL (kWh)** | *Energy Security* | 29,992 | 29,992 | 28,102 | 27,171 | 8,274 | 7,188 | **22,804** |

## 5.2 DRL Flexible System Evolution Strategy

The FD produced via the proposed DRL based approach suggests the development of a highly dynamic decision-making policy, with capacity evolution based on uncertainty realisations in each scenario. The modular nature of the adaptability strategy is graphically captured in Figure 11 below, showing installed nominal capacities over time for different technological components. The plot suggests that in early years the DRL policy presents limited variability, and largely mirrors the findings from the inflexible baselines, with deployment of PV and BESS. Over time, however, it grows increasingly dynamic as more RES capacity is installed within the energy system. This is primarily driven by expansion of wind technologies after about year 7, which then also allows for more EH capacity to be sustainably and reliably operated. Furthermore, in line with near term projections likely to be more accurate, the policy grows more variable over the project life, adapting to the magnitude of uncertainty encountered.

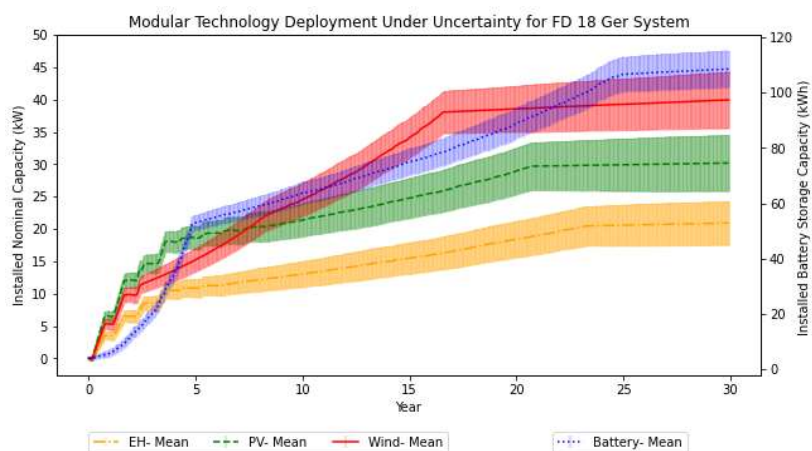

*Figure 11: Cumulative System Capacity Expansion averaged over 2000 episodes showing 95% confidence interval.*

The planning strategy obtained can be further decomposed for improved interpretability by evaluating actions selected under different conditions. Figure 11, for instance, could be broken down into the expansion timing distribution for each technology, and used to recommend options for exercising flexibility at various stages of the project. Other metrics can be developed for this purpose, through a mix of intuition, domain expertise, and trial and error. The ratio of installed renewables to EH nominal capacity provides such an example, as a pattern can be observed where this ratio increases at first, then starts to decrease around year 12 and stabilizes at **around 4.2 to 5.1** in the final periods. The behaviour of the DRL agent under different shortage levels also provides some interesting insights, as shown in Figure 12. It indicates that it is worthwhile to prioritise RES expansion (and wind in particular) under high shortage levels, although important to achieve a balanced generation mix over time. The results also imply that while the majority of EH expansion should happen when there is no unmet load, some capacity should be deployed even in early high shortage years. Most likely, this is because there could still be periods of high EH utilization rate within those years, which can lead to significant cost savings. Energy storage, on the other hand, seems to be prioritized in years with low unmet load levels. This suggests the DRL agent estimates BESS to be particularly helpful at smoothing out those smaller power imbalances, created by intermittent RES generation or stochasticity introduced into the load profiles.

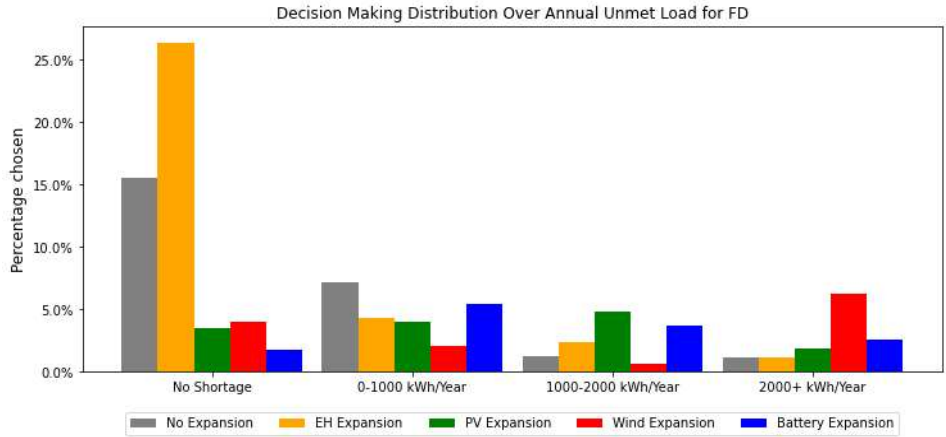

*Figure 12: Distribution of capacity expansion decisions vs shortage levels over 2000 episodes for the DRL agent*

## 5.3 Relative Performance Evaluation
### 5.3.1 Expected Energy System Cost
The low-income communities targeted by these studies motivate prioritising cost as an initial objective. FD achieves significant improvements in this front particularly as compared to BD1 and BD2, reducing ENPC by 25.3% and 20.7%, respectively. The PP enabled inflexible solutions BD3 and BD4 perform slightly better due to the benefits of interconnection, although FD still results in an ENPC 23.3% lower than BD3 and a 15.1% reduction from BD4. Expected cost reductions compared to HOMER deterministic design are even more striking at around 41%, although highly dependent on the economic value of energy not served $c_{ul}$ estimated earlier. Reduction in energy system cost in any given scenario are achieved by FD through an improved adaptability strategy under uncertainty. More practically, this results in decreased expenditure on coal over time, reduced charges due to unmet load, and discounting of capital costs through modular deployment of components, as shown in Figure 11. The greatest economic VoF is recorded for VaR, with FD achieving over a 30% reduction in downside costs as compared to alternatives investigated, although significant improvements are also found in terms of upside potential (VaG). Furthermore, as the capacity expansion strategy developed is highly dynamic and dependant on different realisations of uncertain variables, FD performs much more consistently and is stochastically dominant across *nearly* all scenarios evaluated. The VoF estimated for each economic metric is reported in Table 2.

### 5.3.2 Lifetime Emissions
Even when formulated for purely cost minimisation objectives, the proposed approach is found to yield significant improvements in system design sustainability, as captured by Figure 13. Reductions in estimated lifetime emissions of 54.6%, 67.5%, 52.9% and 56.6% are obtained compared to BD1 and BD2, BD3 and BD4, respectively. Interestingly, the static designs evaluating EH (BD2 and BD4) yield higher net emissions than the others, driven by the UB grid carbon footprint, in line with previous work [80]. This suggests a sustainable ger energy system solution must not rely on the national infrastructure available, but rather on the development of a more distributed and adaptable network, particularly to integrate the multi-energy boundaries encountered. The respective contributions of coal and EH to meeting thermal energy demand for FD over time, shown in Figure 13, suggest that most annual carbon footprint reductions can be achieved around year 18. Nonetheless, some emissions are maintained for the entire project duration, as supplying heat with coal or electricity from the UB grid in certain timesteps can reduce costs. The VoF in the context of sustainability is, in essence, the expected avoided 30-year emissions through the FD as compared to the any of rigid alternatives investigated, estimated at 5,532 tonnes $CO_2$ eq. for this system.

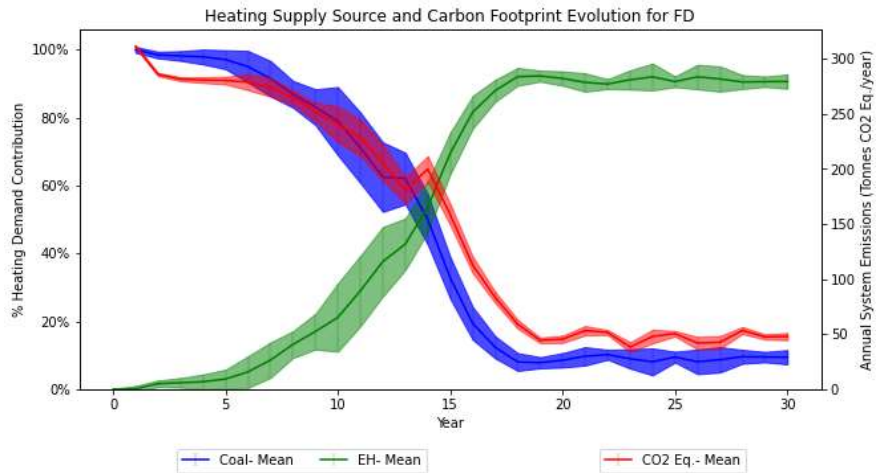

*Figure 13: Heating supply source (left-axis) vs carbon footprint evolution (right-axis) for FD system over 2000 simulated scenarios showing the 95% confidence interval.*

### 5.3.3 Energy Security and Resilience

Outside of improved economic performance and sustainability, FD also achieves substantially enhanced energy security, with reductions in expected TUL of 76.1% compared to BD1 and BD2, and 74.4% and 73.5% from the TUL values obtained in BD3 and BD4. This is driven by the DRL based flexibility strategy, which can closely match capacity and demand over time for the required provision of energy services. The increased generation capacity for FD also reduces reliance by gers on the UB grid, which can be highly unstable due to frequent load shedding. The PP interconnection allows for the smoothing out of load/generation imbalances as compared to the independent nanogrid units, helping increase system resilience for all stakeholders involved, while also reducing BESS requirements. The VoF for energy security is estimated at 22,804 kWh on average for the duration of the project. Intuitively, this implies that for a simulated 30-year horizon, 22,804 kWh of unmet load could be expected to be avoided by using the FD as opposed any of the alternatives considered. Noting that while some unmet load is still recorded for FD, this results from the allowable capacity shortage presented in 2.4 and the intrinsic stochasticity of the environment investigated, as it would require substantial oversizing of components to consistently meet all loads.

### 5.3.4 Local Context Constraints

Average energy system expenses (excluding unmet load charges) are around 83 USD month$^{-1}$ ger$^{-1}$ for FD, and substantially lower for BD1 and BD2 at around 52 and 46 USD month$^{-1}$ ger$^{-1}$, although with significant variation across uncertainty scenarios. The number of timesteps where the budgetary constraints are violated, however, are significantly lower on average for FD. On one hand, this is driven by distributing required investments over project life as compared to significant upfront expenses (and associated replacement costs) recorded for baseline designs. As coal expenses are reduced over time, the budget for expansion decisions for FD increases, which presents nice synergies with the projected load growth, allowing for the deployment of generating technologies only *if and when* needed. Moreover, the PP enabled FD reduces reliance on BESS, which is both the costliest and heaviest technology considered, helping to limit total system mass.

Accordingly, energy system weight for FD is found to be on average around 15% and 19% lower than BD1 and BD2 for the first half of the project, respectively. The much higher final generating capacities installed for FD and extra mass associated with cabling, however, lead to a net increase in end of project system weight of roughly 28% and 22% as compared to BD1 and BD2. BD3/4 usually result in slightly higher weights than BD1/2, but still within the acceptable bounds. Notably, energy system evolution and expansion possibilities over time are a result of reduced coal mass requirements via EH substitution, similar as what is observed for budget. Given the very large amounts of coal consumed by a typical ger, and the availability of regional lightweight systems, mobility constraint violation is thus very unlikely in FD. The few recorded instances can be attributed to energy generation and demand uncertainty as well as potential load shedding, which can result in additional coal mass requirements from that originally estimated.

## 5.4 Value of Plug and Play (VoPP) Operation

In this section, the VoPP is decomposed from the VoF reported in Table 2 to help assess its economic feasibility, according to (39). The results shown in Table 3 suggest it is a worthwhile investment in all cases. These values may be intuitively thought of as the difference in total cost between the same energy system configurations if integrating PP operation or not. FD benefits the most from PP, followed by BD2/4, as the explicit consideration of EH increases system load substantially, and thus motivation for interconnection compared to BD1/3. Nonetheless, for the targeted low-income communities, VoPP for BD1/3 is still significant and can likely be further enhanced. These findings highlight the crucial role of PP in enabling economically feasible mobile energy system solutions, with cost reductions driven by a combination of reduced storage requirements, unmet load penalty and electricity purchased from grid. At the same time, they suggest that a systems approach and flexibility strategy under uncertainty are essential to unlock its full potential, with significantly greater VoPP for FD recorded across all objectives evaluated.

*Table 3: Estimated VoPP and externalities on energy security and sustainability for different system design alternatives*

| Design(s) | VoPP (USD) | % Δ TUL from PP | % Δ $CO_2$ from PP |
|---|---|---|---|
| FD/FDN | **45,313** | -15.1 | -7.3 |
| BD1/3 | **11,134** | -6.3 | -3.2 |
| BD2/4 | **27,312** | -9.4 | -5.6 |

It is important to note, however, that this estimated value is highly dependent on the assumptions used, particularly in terms of herder migration modelling. Significantly greater spatial dispersions than the ones investigated could lead to unsustainable system losses from low voltage distribution, the costs of which may outweigh the benefits generated by PP. To further investigate the effect of this, average distribution losses are plotted against the range of potential cluster radiuses encountered in Figure 14. The "break-even" *starting* cluster radius (where VoPP = 0) is estimated at roughly **12.1 km** for the 18-ger system FD through the economic value of unmet load $c_{ul}$. This implies that in regions with greater spatial dispersions among herders, higher distribution voltages should be considered to maintain an acceptable network efficiency. Alternatively, government policy programs or decision support tools could be developed to incentivize nomadic households to settle closer to each other, helping to capture the full benefits enabled by PP.

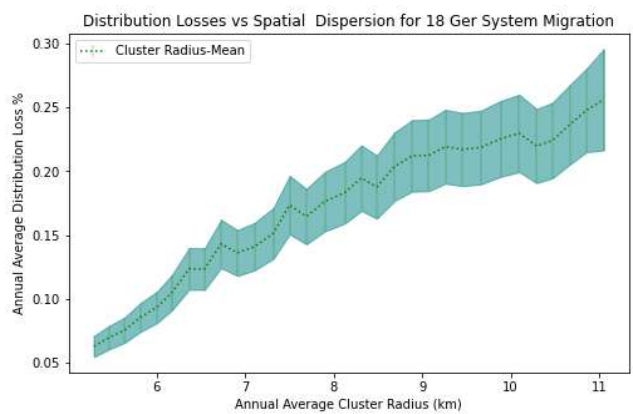

*Figure 14: Annual distribution loss % as a function of herder PP microgrid spatial dispersion*

# 6 Discussion

## 6.1 Changing Economies of Scale

Flexibility in Design is generally motivated by tensions between EoS and modular adaptable deployment, where each can be advantageous depending on the setting. The results presented in the former section, however, suggest that the flexible solution is stochastically dominant and outperforms the baseline designs in nearly all scenarios evaluated. This may be due to the relatively low magnitude of EoS assumed in the cost calculations. Coal expenses, even if significantly reduced, remain the strongest driver of ENPC for nomadic energy systems in Mongolia. While evolving costs for other technologies may influence this finding, an important advantage of the proposed DRL approach is precisely the simplicity to retrain the agent with updated assumptions as they are verified over time. As such, if governmental programs are developed which influence the optimal solution (i.e. subsidy schemes, carbon credits, etc.), they could be quickly integrated into the model to give up to date recommendations. Finally, it is worth noting that removing the budgetary constraints formulated could lead to the discovery of better performing solutions which take advantage of both EoS and flexibility. While this constraint was relaxed in the zero-stage capital expense decision, there are likely to be some practical, logistical and/or economic benefits to ramping up and concentrating investment in certain time periods.

## 6.2 Decarbonisation Pathways

The trade-offs between costs and emissions for inflexible ger energy systems not considering PP were further evaluated through the Non-dominated Sorting Genetic Algorithm (NSGA-II) to the optimisation of BD1 and BD2 [143]. The pareto frontier obtained suggests that high EH capacities should be deployed even by inflexible nanogrid units seeking to minimise ENPC, as they help reduce coal expenses over time - instead purchasing heavily discounted (particularly at night) electricity from the UB grid. The associated carbon footprint, however, means those solutions are estimated to lead to a net increase in expected system emissions over the project horizon. The results presented in the former sections, in fact, suggest that an economically feasible decarbonisation of ger community energy systems is possible even when accounting for the associated uncertainties, due to several important synergies. On one hand, ger communities present substantially higher demand for electricity in the summer months compared to the winter ones, arising from increased refrigerator energy requirements. This allows RES generation to be utilized almost exclusively for EH in the winter months. The relatively high capital expense of RES and EH for low-income communities, as well cultural value associated with the coal stove, nonetheless, remain important limiting factors to the rate of future feasible decarbonisation in Mongolia.

## 6.3 PP Microgrid Feasibility

While this work was focused on the planning rather than operation side, the new control algorithm developments remain largely untested on the field. This is an important issue to be addressed, particularly in terms of overall resulting network efficiency and feasibility at different levels of spatial dispersion. In this study, a ring like interconnection was fixed, however this is likely to vary significantly during actual herder migration, which may strongly impact the performance of a design and flexibility strategy. Additionally, even though only active power was considered here, there are substantial safety and reliability issues associated with low-voltage AC distribution over long distances. Nonetheless, the results from this work imply that there is a significant economic value attached to PP operation, thus motivating the need for more in-depth feasibility studies.

## 6.4 DRL Environment Insights for Community Decision-Making

The analysis presented in Section 5 already provides some interpretable guidelines which could be used by herder communities to determine their energy system expansion strategy. To capture the full value and dynamicity of the proposed approach, however, a decision support tool could be implemented. This would be targeted at improving data visualisation, understanding of the policy optimisation performed, and ultimately enhancing acceptability. A mobile app could be developed which helps community members track energy system state, explore locations of other nearby PP clusters if migration is required, and dynamically recommend actions over time. Given the communication infrastructure is already present, this could be achieved at a low cost. Furthermore, users with different budgetary or mobility requirements from those assumed in this work could enter that data directly in the app, helping generate solutions better tailored to each customer.

### 6.5 Limitations and Future Work Opportunities

While this work presents an important starting point to evaluate the planning of future nomadic and mobile energy systems, there are some other limitations which should be noted. Data on load profiles and projections were difficult to find, and several assumptions had to be made to develop the model from Section 2. A higher resolution heat load profile model, perhaps based on outside temperatures in each hour per day, could allow for a more detailed evaluation of different climate change scenarios. The mobility aspect within the energy system simulation is completely novel and the modelling approach implemented hard to validate. Continuing interactions with local energy sector stakeholders and herder communities will be crucial in this regard. Especially valuable insights may be on the flexibility of migration patterns assumed, and willingness to deviate from them to engage in PP operation in different regions. Additionally, the results obtained are highly dependent on realisations of different cost factors over the project life. These were mostly assumed to be invariant in time for simplicity, and for lack of better information, however, are highly likely to differ over the next 30-years. In order to address data limitations at the time of writing, future work ought to investigate reproducibility of solutions when using different stochastic model parameters. Considering the possibility of new bulk procurement programs, it may be interesting to policy makers to perform a more in-depth sensitivity analysis on the impact of EoS in Mongolia. This can suggest what the break-even values may be, or where the rigid designs outperform the flexible ones.

The action space for the DRL agent could be significantly expanded from this implementation to reflect the relevant decision-making possibilities. Multiple technologies could be considered for expansion at once, for instance. Moreover, connection to the UB or a nearby grid could be explicitly modelled, and respective recommendations produced. Several other potential actions were not evaluated here, particularly higher efficiency stoves or improved ger thermal insulation, which could be highly valuable for decarbonisation objectives. Inverter capacity was also not explicitly analysed, thus likely leading to sub-optimal operation at different points of the project. Closer integration of the control and planning aspects of this problem should also be investigated in future work. While the computational expense required may make the problem intractable, it would be valuable to actually simulate PP operation at a finer time resolution over the entire project horizon. Alternatively, a separate DRL agent could be used to develop more advanced operational strategies than the LFS implemented here. Flexible demand response and prioritised load shedding, for instance, are likely to help reduce ENPC incurred by mobile energy systems.

From an algorithmic point of view, there are several more interesting avenues for future research. More specifically, a multi-agent formulation could be developed to better evaluate the costs and benefits to individual herder households, and the resulting optimal flexibility strategies. Recent advancements in distributional DRL allow the formulation of a policy based on the entire distribution of rewards rather than purely expected value [144]. Recommended strategies could thus be tailored to risk-seeking or risk averse decision makers, improving acceptability, and understanding. Finally, exploring alternative DNN architectures which are better tailored to the tackled problem could lead to further improvements. CNNs could be used to map relationship among features which may not be intuitive at first, for instance, while RNNs could help capture the seasonal nature of decision making required in order to better adapt agent policy between winter and summer months.

# 7   Conclusions

In this paper, a DRL-based approach was developed for the design and planning of a mobile energy system supply system, and illustrated through a case study on Mongolian herder communities. Motivated by the potential for increased climate induced migration and increasingly decentralized generation, a novel modelling framework to integrate mobility considerations was developed. This analysis was enabled by recent "Plug and Play" control developments, allowing temporary interconnection among nomadic households and/or the main grid. Contrary to popular energy system planning methods, the data-driven approach implemented allowed for the tractable integration of several sources of uncertainty as well as mobility within the evaluation model. Both heat and electricity were considered, leading to a more holistic evaluation of different solution alternatives than previously available. The design and planning strategy for a highly flexible and modular energy system was thus optimised through an actor-critic algorithm, with a reasonable computational expense. Benchmark inflexible designs with economies of scale were compared across key economic, sustainability and resilience indicators such as Cost, Equivalent Emissions and Total Unmet Load. The results suggest that the proposed approach can lead to economically feasible mobile energy systems, even in the case of budgetary and mobility constraints. They also show that a DRL-based flexible capacity expansion strategy can offer a highly dynamic and adaptable energy system design, with multi-objective improvements compared to rigid baselines available from institutional programs. Additionally, the value generated by "Plug and Play" operation was also estimated using a variation of real-options theory, suggesting it is a worthwhile investment in all cases. The spatial dispersion among herders within a cluster, however, can be an important limiting factor due to the low voltage of distribution. Finally, while this paper provides an important advancement to enabling nomadic energy access, there are several limitations and opportunities for future work, which could enhance the applicability and overall impact of the methods presented.

# 9 Appendix

## 9.1 I – Mathematical Formulation of Power Flow Model

$$P_{NEL}^{n,t} = ED^{n,t} - P_{res}^{n,t} \qquad 41$$

$$P_{res}^{n,t} = P_{wind,adj}^{n,t} + P_{pv,adj}^{n,t} \qquad 42$$

$$P_{Load,Net}^{n,t} = HD^{n,t} + ED^{n,t} - P_{res}^{n,t} \qquad 43$$

$$P_{HD,eh,res}^{n,t} = \min(HD^{n,t}, P_{eh,max}^{n,t}, -P_{ED,Load,Net}^{n,t}) \quad , P_{NEL}^{n,t} \leq 0 \qquad 44$$

$$P_{NHL}^{n,t} = HD^{n,t} - P_{eh,res}^{n,t} \qquad 45$$

$$P_{res,PL}^{n,t} = P_{res}^{n,t} - P_{HD,eh,res}^{n,t} - P_{NEL}^{n,t} \quad , P_{ED,Load,Net}^{n,t}, P_{HD,Load,Net}^{n,t} \leq 0 \qquad 46$$

$$n \in [N_1, , N_2, N_3, .. N_N], \forall n, t \qquad 47$$

$$P_{ba,\ out\ max}^{n,t} = \theta_{ba}^{n,t} C_{out} \qquad 48$$

$$P_{ba,in,max}^{n,t} = \theta_{ba}^{n,t} C_{in} \qquad 49$$

$$P_{ba}^{n,t} = \begin{cases} -\min(P_{ba,out,max}^{n,t}, P_{NEL}^{n,t}) & P_{NEL}^{n,t} \geq 0 \quad (a) \\ -\min(P_{ba,out,max}^{n,t}, P_{HD,Load,Net}^{n,t}) & P_{res,PL}^{n,t}, P_{ED,Load,Net}^{n,t} \leq 0\ ;\ P_{HD,Load,Net}^{n,t} \geq 0\ (b) \\ \min(P_{ba,in,max}^{n,t}, P_{res,PL}^{n,t}) & P_{res,PL}^{n,t} \geq 0\ ;\ P_{HD,Load,Net}^{n,t}, P_{ED,Load,Net}^{n,t} \leq 0 \quad (c) \end{cases} \qquad 50$$

$$E_{ba}^{n,t} = E_{ba}^{n,t-1} + P_{ba}^{n,t}\sqrt{\eta_{ba}} + E_{ba,new}^{n,t} \qquad 51$$

$$E_{ba}^{n,t} = E_{ba}^{n,t-1} + P_{ba}^{n,t}\sqrt{\eta_{ba}} + E_{ba,new}^{n,t} \qquad 52$$

$$P_{ul,PB}^{n,t} = \begin{cases} P_{Load,Net}^{n,t} - (E_{ba}^{n,t} - E_{ba,min}^{n,t}), & P_{Load,Net}^{n,t} \geq 0 \\ 0 & otherwise \end{cases} \qquad 53$$

$$P_{HD,Load,Net,PB}^{n,t} = HD^{n,t} - P_{eh,res}^{n,t} - P_{eh,batt}^{n,t} \tag{54}$$

$$P_{HD,Load,Net,PB,ES}^{t} = \sum_{n \in N} P_{HD,Load,Net,PB}^{n,t} \tag{55}$$

$$P_{eg}^{n,t} = \min(P_{res,PL}^{n,t} - P_{ba}^{n,t}, 0) \qquad P_{ba}^{n,t} \geq 0 \tag{56}$$

$$n \in N_{deficit} : P_{eg}^{n,t} \leq 0 \tag{57}$$

$$n \in N_{surplus} : P_{eg}^{n,t} > 0 \tag{58}$$

$$P_{ul,es}^{t} = \sum_{n \in N_{deficit}} P_{ul,PB}^{n,t} + P_{HD,Load,Net,PB}^{n,t} \tag{59}$$

$$P_{uel,es,bpp}^{t} = \sum_{n \in N_{deficit}} P_{ul,PB}^{n,t} \tag{60}$$

$$P_{eg,es}^{t} = \sum_{n \in N_{surplus}} P_{eg}^{n,t} \tag{61}$$

$$P_{dist,es}^{t} = \min(P_{ul,es}^{t}, P_{eg,es}^{t}, \theta_{PPmg}^{t}, P_{dist,es,max}^{t}) \tag{62}$$

$$s.t. \; 0 \leq P_{dist,es}^{t} \leq 0.3 \, (ED_{es}^{t}) \tag{63}$$

$$L_{dist}^{t,n} = \frac{L_{cb}^{t}}{N} \tag{64}$$

$$P_{dist}^{t,n} = \frac{P_{dist,es}^{t}}{N} \tag{65}$$

$$R_{dist,ref}^{t,n} = \frac{\rho_{cb} L_{dist}^{t,n}}{A_{cs}^{cb}} \tag{66}$$

$$R_{dist,actual}^{t,n} = R_{dist,ref}^{t,n}[1 + \alpha_{copper}(T_{actual} - T_{ref})] \tag{67}$$

$$P_{dl,es}^{t} = \sum_{n \in N} 3 R_{dist,actual}^{t,n} {I_{dist}^{t,n}}^{2} \tag{68}$$

$$I_{dist}^{t,n} = \frac{P_{dist}^{t,n}}{V_{dist}^{t,n}} \tag{69}$$

$$P_{res,exc,pp}^t = P_{eg,es}^t - P_{dist,es}^t \qquad (70)$$

$$P_{uel,es,ppp}^t = P_{uel,es,bpp}^t - (P_{dist,es}^t - P_{dl,es}^t) \qquad (71)$$

$$DL\ (\%) = 100\frac{P_{dl,es}^t}{P_{dist,es}^t} \qquad (72)$$

## 9.2 Model Parameters and Variables Values

Please refer to the nomenclature table presented at the beginning of the manuscript for the definition and respective reference for each of these variables/ parameters.

**Technical Parameters and Values/Ranges**

| | | | |
|---|---|---|---|
| $T$ | 30 years / 360 months | $K_{eh}$ | 1.52 USD/W |
| $\lambda$ | 5% | $K_{ba}$ | 0.7 USD/ Wh |
| $\eta_{inv}$ | 90% | $\alpha_{MS1}$ | 0.95 |
| $\eta_{batt}$ | 75% | $\alpha_{MS2}$ | 0.85 |
| $\eta_{coal\ stove}$ | 25% | $\varepsilon_{pe,avg}$ | 0.769 |
| $\mu_{coal}$ | 14.6 MJ/kg | $k^{salv}$ | 0.7 |
| $A_{cs,cb}$ | 1.5 mm² | $L_{pv}^{max}$ | 25 years |
| $\rho_{cb}$ | 1.678x10⁻⁸ Ωm | $L_{wind}^{max}$ | 20 years |
| $T_{ref}$ | 20 °C | $L_{pe}^{max}$ | 15 years |
| $\alpha_{cb,adj}$ | 0.393 % / °C | $L_{eh}^{max}$ | 13 years |
| $EF_{grid}$ | 0.711 tCO₂ eq./ MWh | $c_{coal}$ | 40 USD/ t |
| $EF_{coal}$ | 1.37 tCO₂ eq./ 1000 kg | $c_{om,pv}$ | 0.0048 $/kW/month |
| $\theta_{UB\ Grid}$ | 2 kW | $c_{om,wind}$ | 0.0032 $/kW/month |
| $A_{grid}$ | 1 | $c_{om,pe}$ | 0.0001 $/kW/month |
| $B_{grid}$ | 9 | $c_{ul}$ | 0.3417 USD/kWh |
| $c_{cc}$ | 0.04 USD/ kWh | $c_{grid}$ | 0.041 USD/kWh |
| $c_{ic}$ | 5.2 USD/m | $p_{grid}$ | 0.17 USD/kWh |
| $c_{pe,avg}$ | 0.352 USD/W | | |
| $K_{pv}$ | 2.64 USD/W | | |
| $K_{wind}$ | 1.91 USD/W | | |

**GBM Model(s) Parameters**

| | |
|---|---|
| $\mu_{migration}$ | 0.5% |
| $\sigma_{migration}$ | 5% |
| $\mu_{ED}$ | 3.14% |
| $\sigma_{ED}$ | 15% |
| $\mu_{HD}$ | 1% |
| $\sigma_{HD}$ | 6.5% |
| $\sigma_{PV}$ | 19.3% |
| $\sigma_{wind}$ | 14.2% |